\documentclass[12pt,reprint,aip,jcp,floatfix]{revtex4-2}
\usepackage[english]{babel}

\usepackage{graphicx}
\usepackage{dcolumn}

\usepackage[fleqn]{amsmath}
\usepackage{amssymb}
\usepackage{bm}
\usepackage{subcaption}
\usepackage{threeparttable}
\usepackage{bigints}
\usepackage{rotating}
\usepackage{siunitx}
\usepackage[colorlinks=true,linkcolor=blue]{hyperref}
\usepackage{comment}
\usepackage{algorithm}
\usepackage{algpseudocode}

\makeatletter
\let\OldStatex\Statex
\renewcommand{\Statex}[1][3]{%
  \setlength\@tempdima{\algorithmicindent}%
  \OldStatex\hskip\dimexpr#1\@tempdima\relax}
\makeatother

\usepackage{framed}

\usepackage{verbatim}
\usepackage[usenames]{color}
\definecolor{cfb}{rgb}{0.0,0.9,0.9}
\definecolor{orange}{rgb}{1.0,0.45,0.0}
\definecolor{lila}{rgb}{0.7,0.0,1.0}

\newcommand{\mycm}[1]{\textsf{/* #1 */}}

\renewcommand{\vec}[1]{\boldsymbol{#1}}

\newcommand{\vect}[1]{\boldsymbol{#1}}

\newcommand{\dV}[1]{d^3\vect{#1}}

\begin{document}


\title{Exact two-component Hamiltonians for relativistic quantum chemistry: \\
       Two-electron picture-change corrections made simple}

\author{Stefan Knecht}
\email{s.knecht@gsi.de}
\affiliation{SHE Chemie, GSI Helmholtzzentrum für Schwerionenforschung GmbH,
Planckstr. 1, 64291 Darmstadt, Germany}
\affiliation{Department of Chemistry, Johannes-Gutenberg Universit{\"a}t Mainz, Duesbergweg 10-14, 55128 Mainz, Germany}

\author{Michal Repisky}
\email{michal.repisky@uit.no}
\affiliation{Hylleraas Centre for Quantum Molecular Sciences, University of Troms\o, N-9037 Troms\o, Norway}

\author{Hans J{\o}rgen Aagaard Jensen}
\affiliation{Department of Physics, Chemistry and Pharmacy, University of Southern Denmark, Campusvej 55, DK-5230 Odense M, Denmark}

\author{Trond Saue}
\affiliation{Laboratoire de Chimie et Physique Quantiques (CNRS UMR 5626),
Universit\'e Toulouse III -- Paul Sabatier, 118 Route de Narbonne, F-31062 Toulouse cedex, France}

\begin{abstract}
Based on self-consistent field (SCF) atomic mean-field (amf) quantities, we present two simple, yet computationally efficient and numerically 
accurate matrix-algebraic approaches to correct both scalar-relativistic \textit{and} spin-orbit two-electron picture-change effects (PCE) arising within  an exact two-component (X2C) Hamiltonian framework. Both approaches, dubbed amfX2C and e(xtended)amfX2C, allow us to uniquely tailor PCE corrections to mean-field models, \textit{viz.}\ Hartree--Fock or Kohn--Sham DFT, in the latter case also avoiding the need of a point-wise calculation of exchange--correlation PCE corrections. We assess the numerical performance of these PCE correction models on  spinor energies of group-18 (closed-shell) and group-16 (open-shell) diatomic molecules, achieving a consistent $\approx\!10^{-5}$ Hartree accuracy compared to reference four-component data. Additional tests include SCF calculations of molecular properties such as absolute contact density and contact density shifts in copernicium fluoride compounds (CnF$_{n}$, n=2,4,6), as well as equation-of-motion coupled cluster calculations of X-ray core ionization energies of 
$5d$ and $6d$-containing molecules, where we observe an excellent agreement with reference data. To conclude, we are confident that our (e)amfX2C PCE correction models constitute a fundamental milestone towards a universal and reliable relativistic two-component quantum chemical approach, maintaining the accuracy of the parent four-component one at a fraction of its computational cost.
\end{abstract}

\date{\today}


\maketitle

\onecolumngrid

\section{Introduction}\label{sec:intro}

Advancing with rapid strides in the past decades, relativistic quantum chemical approaches are becoming a standard ingredient in the computational toolbox of theoretical chemists. Notwithstanding important steps forward to turn a fully relativistic quantum-chemical approach based on the four-component Dirac formalism into a handy tool \cite{dyall_faegri_qc,reiher_qc,saue11a,liuw14,dirac-paper,ReSpect2020,BAGEL2018,BDF2020,bertha2020}, much of the success is due to the fast-paced development and implementation of efficient quasi-relativistic ``exact" two-component approaches (X2C)~\cite{x2c} in various originally non-relativistic popular quantum-chemistry software packages within the past two decades.
This has become possible by making use of a matrix-algebra formalism rather than setting out from an (order-by-order) based operator  formalism~\cite{doug74,hess86,chan86a,lent93a,lent96a,bary02a,reih04a,reih04b,reih06}.

In relativistic quantum chemistry, the common starting point for almost all of the matrix-algebra-based two-component (2c) Hamiltonian approaches, whether formulated within an elimination ansatz~\cite{dyal97a,fila04a,fila06a,fila07a} or in a unitary-decoupling framework,\cite{hjaaj:rehetalk,kutz05,kutz06b,liuw06b,ilias07,liuw09a,kone16a} has been the four-component (4c) \textit{one-electron} Dirac Hamiltonian in the electrostatic potential of fixed nuclei.\cite{saue11a}
We will in the following refer to the 4c Hamiltonian used to construct a 2c model as the \textit{defining} 4c Hamiltonian. 
 In the case of the one-electron X2C Hamiltonian scheme (1eX2C), the two-electron (2e) interaction term is omitted from the defining 4c Dirac Hamiltonian. Consequently, the resulting 2c Hamiltonian is to be considered ``exact" only wrt the inclusion of 1e terms in the defining 4c Hamiltonian \cite{peng12a}, while the account of the 2e interaction is postponed to \textit{after} having carried out the unitary decoupling of the 1e Hamiltonian and the ensuing restriction to the upper (``electrons-only") 2c spinor basis. 
 Such an approach usually implies the use of the \textit{untransformed} 2e interaction term in the 1eX2C basis set, giving rise to 2e picture-change effects (2ePCEs).
A noticeable exception exists, though, and has been coined the molecular--mean field exact two-component approach (mmfX2C).~\cite{sikkema09} 
In contrast to the 1eX2C scheme, the mmfX2C \textit{ansatz} is based on a unitary decoupling of the 4c molecular mean-field \textit{Fock matrix} after having converged the 4c self-consistent field (SCF) Hartree–Fock equations. 
Although strictly matching with the SCF results of those obtained with the corresponding defining 4c Hamiltonian,~\cite{liuw06b,sikkema09} the mmfX2C approach will still be an approximation in any ensuing post-SCF electron-correlation step for which the untransformed 2e interaction term replaces its complete (transformed) counterpart.

Hence, the extent to which 2ePCEs are accounted for in an X2C Hamiltonian based relativistic quantum-chemical framework is essential for its applicability to address the electronic-structure theory problem in many-electron (molecular) systems involving elements across the entire periodic table.~\cite{peng12a} 
To this end, we note  that the 2e interaction term can be decomposed into a spin-free or scalar-relativistic (SC) as well as  a spin-dependent or spin-orbit (SO) part,~\cite{dyall94,dyall_faegri_qc} where both the two-electron scalar-relativistic (2eSC) and two-electron spin-orbit (2eSO) terms serve as a screening of their 1e counterparts. Whereas much attention has been paid in the past to efficiently take into account 2eSO PCEs based on a variety of \textit{ans\"atze}, the 2eSC contributions are, curiously, less commonly included in correction schemes for 2ePCEs as has been comprehensively summarized in the Introduction of Ref.~\citenum{liuj18}. Examples of approximate 2eSO corrections range from using 
(i) a parametrized model approach based on nuclear charges multiplied with element and angular-momentum specific screening factors in the evaluation of 1eSO integrals;~\cite{blum62a,blum63a} 
(ii) a mean-field SO approach~\cite{hess96} which has been the basis for the widely popular~\textsc{AMFI} module \cite{schimmel96}\ interfaced for example with the software packages \textsc{DIRAC},\cite{dirac-paper} \textsc{OpenMolcas},\cite{openm19a}, and \textsc{DALTON};\cite{daltonpaper} 
(iii)\ an approach that exploits atomic model densities obtained within the framework of Kohn--Sham DFT (KS-DFT).\cite{vanW05a,peng07b,auts12b} 
Interestingly, although the latter model-density based correction schemes are rare examples which in addition to corrections for 2eSO PCEs do provide corrections for 2eSC PCEs, the resulting correction terms do not discriminate between the use of {different} exchange-correlation functionals employed in a \textit{molecular} X2C-Hamiltonian based Kohn-Sham DFT calculation.
The screening factors of type (i) are sometimes referred to as "Boettger factors". 
In current usage they have  been obtained for a second-order, truncated 2c Hamiltonian \textit{ansatz} (i.e. second-order Douglas--Kroll--Hess (DKH2)) within the framework of density functional theory (DFT) \cite{boet00a} but are, remarkably, also commonly employed in X2C-Hamiltonian based wave function theory (WFT) approaches~\cite{lixi19a,lixi19b,lixi20a} --

In their most recent work on suitable 2ePCE corrections for the X2C Hamiltonian, Liu and Cheng~\cite{liuj18}\ proposed an atomic mean-field (amf) approach which exploits a mean-field approximation for PCEs originating from the 2eSO contribution, dubbed SOX2CAMF by them, and combines ``the four main ideas in relativistic quantum chemistry (\ldots): the X2C decoupling scheme, the 1e approximation for SC effects (i.e., the neglect of the scalar 2e picture-change effects), the mean-field SO approach, and the atomic approximation for the 2eSO interactions".~\cite{liuj18} Thus, a key feature of the SOX2CAMF model is that it does not require the evaluation of any \textit{molecular} relativistic 2e integrals. Although it has in the meantime been employed successfully in highly sophisticated electron correlation calculations of heavy-element complexes,\cite{liuj21a} limitations of the underlying atomic approximation to account for  2eSO PCEs have recently been pointed out in the context of zero-field splittings of first-row transition metal complexes.~\cite{netz21a} 

In this paper we introduce an atomic mean-field (amfX2C) as well as an extended atomic mean-field (eamfX2C) approach within the X2C Hamiltonian framework which not only takes into account the above mentioned four main ideas in relativistic quantum chemistry but also amends them such that the resulting amfX2C and eamfX2C approaches will bridge the gap between a full molecular 4c and mmfX2C framework in a computationally efficient, yet highly accurate way. In contrast to most existing correction schemes for 2ePCE, our amfX2C and eamfX2C approaches are laid out to comprise \textit{full} 2ePCE corrections, that is treating the 2eSO \textit{and} 2eSC ones on the same footing, whether they arise from the (relativistic) 2e Coulomb, Coulomb-Gaunt, or Coulomb-Breit interaction. Moreover, our \textit{ansatz} takes into account the characteristics of the underlying correlation framework, \textit{viz.}, WFT or (KS-)DFT, which enables us to introduce tailor-made exchange-correlation-specific corrections for 2ePCEs. Setting out from the idea of an amf approach within the amfX2C Hamiltonian model -- formulated for a WFT-based HF and a DFT framework in Sections \ref{sec:theory:HF}\ and \ref{sec:theory:KS}, respectively -- the \textit{extended} amfX2C approach encompasses two-center 2e contributions obtained in a molecular framework. The implications arising from the resulting eamfX2C approach, including its potential shortcomings and particular advantages are then discussed in Section \ref{sec:eamf-and-remarks}. 
The numerical accuracy of both (e)amfX2C Hamiltonian models are assessed based on the calculation of a variety of valence and core-like molecular properties in Section \ref{sec:results} where the computational details are given in Section \ref{sec:compdet}. We summarize our results and findings in Section \ref{sec:conclu}\ and provide a prospect of future developments. 

\section{Theory}\label{sec:theory}

\subsection{The amfX2C Hamiltonian -- Hartree--Fock framework}\label{sec:theory:HF}

A convenient starting point for our derivations to arrive at suitable corrections for 2ePCEs in an X2C Hamiltonian framework is to consider the closed-shell 4c HF equations  based on the Dirac--Coulomb Hamiltonian. Without any loss of generality, we may consider these equations in orthonormal basis
\begin{equation} \label{eq:HFeq}
   \vec{F}^{\text{4c}} \vec{c}^{\text{4c}}_{i}
   =
   \vec{c}^{\text{4c}}_{i} \epsilon^{\text{4c}}_{i},
\end{equation}
also because our computer implementations generate corresponding 2c quantities in such a basis.\cite{ilias07,kone16a} 
The HF energy and the Fock matrix have their usual definitions
\begin{align}
   E^{\text{4c}}
   &=
   E^{\text{4c,1e}} + E^{\text{4c,2e}}
   =
   \displaystyle\sum_{\mu\nu} h^{\text{4c}}_{\mu\nu} D^{\text{4c}}_{\nu\mu}
   + 
   \frac{1}{2}\sum_{\mu\nu\kappa\lambda} D^{\text{4c}}_{\nu\mu} G^{\text{4c}}_{\mu\nu,\kappa\lambda} D^{\text{4c}}_{\lambda\kappa}
   \label{eq:HFenergy}
   \\
   F_{\mu\nu}^{\text{4c}}
   &=
   F_{\mu\nu}^{\text{4c,1e}} + F_{\mu\nu}^{\text{4c,2e}}
   =
   h^{\text{4c}}_{\mu\nu}
   +\displaystyle
   \sum_{\kappa\lambda}
   G^{\text{4c}}_{\mu\nu,\kappa\lambda} D^{\text{4c}}_{\lambda\kappa}
   =
   \frac{\text{d}E^{\text{4c}}}{\text{d}D^{\text{4c}}_{\nu\mu}}
   \label{eq:Fock_matrix}
\end{align}
in terms of the atomic orbital (AO) density matrix
\begin{equation}
  D^{\text{4c}}_{\mu\nu}=\sum_{i}^{\text{occ}}c^{\text{4c}}_{\mu i}c^{\text{4c}^{\ast}}_{\nu i}
\end{equation}  
and the matrix of anti-symmetrized two-electron AO integrals
\begin{equation} \label{eq:eri}
   G^{\text{4c}}_{\mu\nu,\kappa\lambda}
    =
   \mathcal{I}^{\text{4c}}_{\mu\nu,\kappa\lambda}
   -
   \mathcal{I}^{\text{4c}}_{\mu\lambda,\kappa\nu}
   ;\quad
   \mathcal{I}^{\text{4c}}_{\mu\nu,\kappa\lambda}
   \equiv
   \iint
   \Omega_{\mu\nu}^{\text{4c}}(\vec{r}_{1})
   r_{12}^{-1}
   \Omega_{\kappa\lambda}^{\text{4c}}(\vec{r}_{2})
   d^{3}\vec{r}_{1}d^{3}\vec{r}_{2},
\end{equation}
the latter expressed in terms of overlap distribution functions~\cite{ReSpect2020}
\begin{equation} \label{eq:omega0}
  \Omega_{\mu\nu}^{\text{4c}}(\vec{r})
   \equiv 
   \int \chi_{\mu}^{\dagger}(\vec{r}^{\prime})\delta^3(\vec{r}^{\prime}-\vec{r})\chi_{\nu}(\vec{r}^{\prime})d^3\vec{r}^{\prime}
   =  
   \chi_{\mu}^{\dagger}(\vec{r})\chi_{\nu}(\vec{r})
\end{equation}
over 2-component basis functions $\chi_{\mu}(\vec{r})$;  formally the basis functions are 4-component objects, but with the lower or upper two components zero according to whether they are large (L) or small (S). 

The \textit{converged} HF equations, Eq.~\eqref{eq:HFeq}, form the starting point for the mmfX2C approach,\cite{sikkema09} where the Fock matrix and corresponding positive-energy molecular-orbital (MO)  coefficients $(+)$ are picture-changed to 2c form
\begin{equation} \label{eq:X2Cmmf}
   \tilde{F}^{\text{2c}}_{\mu\nu}
   \equiv
   \Big[
       \vec{U}^{\dagger} \vec{F}^{\text{4c}} \vec{U}
   \Big]^{\text{LL}}_{\mu\nu}
   ;\qquad
   \tilde{c}^{\text{2c}}_{\mu i}
   \equiv
   \Big[ 
        \vec{U}^{\dagger} \vec{c}^{\text{4c}}
   \Big]^{\text{L}+}_{\mu i}
\end{equation}
(note that we use tildes to indicate picture-change transformed quantities). 
These quantities, together with the anti-symmetrized two-electron AO integrals, Eq.~\eqref{eq:eri}, are then used to build the normal-ordered Hamiltonian for use in subsequent wave-function based correlation methods. 

In the present case we rather seek to carry out the SCF-iterations themselves in 2c mode, but in a manner such that we optimally reproduce the 4c results. A first important observation comes from consideration of the picture-change transformed Fock matrix
\begin{equation}\label{eq:Fpc}
  \tilde{F}^{\text{2c}}_{\mu\nu}
  =
   \sum_{XY} \sum_{\alpha\beta}
   \big[ U^{\dagger}   \big]^{\text{LX}}_{\mu\alpha}
   \big[ F^{\text{4c}} \big]^{\text{XY}}_{\alpha\beta}
   \big[ U             \big]^{\text{YL}}_{\beta\nu}
   ;\quad \text{X,Y} \in \text{L, S}
\end{equation}
Noting that the positive-energy 4c MO-coefficients can be expressed in terms of their 2c counterparts
\begin{equation} \label{eq:4cinv}
   \vec{c}^{\text{4c};+}
   =
   \vec{U} \tilde{\vec{c}}^{\text{2c}}
   \qquad\Rightarrow\qquad
   \big[ c^{\text{4c}} \big]^{\text{X}+}_{\mu i}
   =
   \sum_{\nu}
   \big[ U \big]^{\text{XL}}_{\mu\nu} 
   \big[ \tilde{c}^{\text{2c}} \big]_{\nu i}
   ;\qquad
   \text{X} \in \text{L, S},
\end{equation}
we can reformulate the two-electron 2c Fock matrix as
\begin{align} \label{}
   \tilde{F}^{\text{2c,2e}}_{\mu\nu}
   & =
   \sum_{XY} \sum_{\alpha\beta}
   \big[ U^{\dagger}      \big]^{\text{LX}}_{\mu\alpha}
   \big[ F^{\text{4c,2e}} \big]^{\text{XY}}_{\alpha\beta}
   \big[ U                \big]^{\text{YL}}_{\beta\nu}
   \\
   & =
   \sum_{XY} \sum_{\alpha\beta}
   \big[ U^{\dagger}   \big]^{\text{LX}}_{\mu\alpha}
   \left\{
   \sum_{UV} \sum_{\gamma\delta}
   \big[ G^{\text{4c}} \big]^{\text{XY,UV}}_{\alpha\beta,\gamma\delta}
   \big[ D^{\text{4c}} \big]^{\text{VU}}_{\delta\gamma}
   \right\}
   \big[ U             \big]^{\text{YL}}_{\beta\nu}
   \nonumber
   \\
   & =
   \sum_{XY} \sum_{\alpha\beta}
   \big[ U^{\dagger}     \big]^{\text{LX}}_{\mu\alpha}
   \left\{
   \sum_{UV}
   \sum_{\gamma\delta}
   \big[ G^{\text{4c}}  \big]^{\text{XY,UV}}_{\alpha\beta,\gamma\delta}
   \sum_{\kappa\lambda}
   \sum_{i}
   \big[ U                         \big]^{\text{VL}}_{\delta\kappa}
   \big[ \tilde{c}^{\text{2c}}     \big]_{\kappa i}
   \big[ \tilde{c}^{\text{2c}^{*}} \big]_{\lambda i}
   \big[ U^{*}                     \big]^{\text{UL}}_{\gamma\lambda}
   \right\}
   \big[ U                         \big]^{\text{YL}}_{\beta\nu}
   \nonumber
   \\
   & =
   \sum_{\kappa\lambda}
   \left\{
   \sum_{XYUV}
   \sum_{\alpha\beta\gamma\delta}
   \big[ U^{\dagger}           \big]^{\text{LX}}_{\mu\alpha}
   \big[ U^{\dagger}           \big]^{\text{LU}}_{\lambda\gamma}
   \big[ G^{\text{4c}}         \big]^{\text{XY,UV}}_{\alpha\beta,\gamma\delta}
   \big[ U                     \big]^{\text{VL}}_{\delta\kappa}
   \big[ U                     \big]^{\text{YL}}_{\beta\nu}
   \right\}
   \big[ \tilde{D}^{\text{2c}} \big]_{\kappa\lambda}
   ;\qquad
   \text{X,Y,U,V} \in \text{L, S}.
\end{align}
As a consequence
\begin{equation} \label{eq:2cFock}
   \tilde{F}^{\text{2c}}_{\mu\nu}
   =
   \tilde{h}^{\text{2c}}_{\mu\nu}
   +
   \sum_{\kappa\lambda}
   \tilde{G}^{\text{2c}}_{\mu\nu,\kappa\lambda}
   \tilde{D}^{\text{2c}}_{\lambda\kappa}.
\end{equation}
We see that the picture-change transformed Fock matrix can be expressed in
terms of the picture-changed transformed coefficients as well as the picture-changed one- and two-electron
integrals. By similar manipulations we can also show that the 4c HF energy can be expressed 
in terms of corresponding 2c quantities, that is
\begin{align}
\label{eq:HF-4c-2c-energy}
   E^{4c}
   &=
   \sum_{XY} \sum_{\mu\nu}
   \big[ h^{\text{4c}} \big]^{\text{XY}}_{\mu\nu}
   \big[ D^{\text{4c}} \big]^{\text{YX}}_{\nu\mu}
   +
   \frac{1}{2}
   \sum_{XYUV} \sum_{\mu\nu\kappa\lambda}
   \big[ D^{\text{4c}} \big]^{\text{YX}}_{\nu\mu}
   \big[ G^{\text{4c}} \big]^{\text{XYUV}}_{\mu\nu,\kappa\lambda}
   \big[ D^{\text{4c}} \big]^{\text{VU}}_{\lambda\kappa}
   \nonumber
   \\[0.2cm]
   &=
   \sum_{\mu\nu}
   \big[ \tilde{h}^{\text{2c}} \big]_{\mu\nu}
   \big[ \tilde{D}^{\text{2c}} \big]_{\nu\mu}
   +
   \frac{1}{2}
   \sum_{\mu\nu\kappa\lambda}
   \big[ \tilde{D}^{\text{2c}} \big]_{\nu\mu}
   \big[ \tilde{G}^{\text{2c}} \big]_{\mu\nu,\kappa\lambda}
   \big[ \tilde{D}^{\text{2c}} \big]_{\lambda\kappa}
   =
   \tilde{E}^{\text{2c,1e}} + \tilde{E}^{\text{2c,2e}}.
\end{align}
We conclude that provided we start from the \emph{correctly transformed} set of integrals the 2c SCF will converge to the coefficients
$\big\{ \tilde{\vec{c}}^{\text{2c}}_{i} \big\}$
corresponding to the converged 4c SCF and we shall furthermore reproduce the positive orbital energies as well as total energy of the parent 4c HF. However, the picture-change transformation $\vec{U}$ associated with the converged 4c Fock matrix is \textit{not} available at the start of the SCF-iterations, forcing us to introduce approximations.

With this in view, a second important observation arises from
comparison of Eq.~\eqref{eq:2cFock} with the Fock matrix built with \textit{untransformed} two-electron integrals
$G^{\text{2c}}_{\mu\nu,\kappa\lambda}$
\begin{equation} \label{eq:2cFock_nopoc}
   F^{\text{2c}}_{\mu\nu}
   =
   \tilde{h}^{\text{2c}}_{\mu\nu}
   +
   \sum_{\kappa\lambda}
   G^{\text{2c}}_{\mu\nu,\kappa\lambda}
   \tilde{D}^{\text{2c}}_{\lambda\kappa}.
\end{equation}
We immediately find that their difference expresses the picture-change
correction of the two-electron integrals
\begin{equation} \label{eq:DeltaF2}
   \Delta\tilde{F}^{\text{2c}}_{\mu\nu}
   =
   \tilde{F}^{\text{2c}}_{\mu\nu}
   -
   F^{\text{2c}}_{\mu\nu}
   =
   \sum_{\kappa\lambda}
   \Delta\tilde{G}^{\text{2c}}_{\mu\nu,\kappa\lambda}
   \tilde{D}^{\text{2c}}_{\lambda\kappa}
   ;\qquad
   \Delta\tilde{G}^{\text{2c}}_{\mu\nu,\kappa\lambda}
   =
   \tilde{G}^{\text{2c}}_{\mu\nu,\kappa\lambda}
   -
   G^{\text{2c}}_{\mu\nu,\kappa\lambda}.
\end{equation}
Moreover, this differential Fock matrix may be used to correct the two-electron HF energy
\begin{align} \label{}
   \tilde{E}^{\text{2c,2e}}
   =
   \frac{1}{2}
   \sum_{\mu\nu\kappa\lambda}
   \tilde{D}^{\text{2c}}_{\nu\mu}
   \tilde{G}^{\text{2c}}_{\mu\nu,\kappa\lambda}
   \tilde{D}^{\text{2c}}_{\lambda\kappa}
   =
   \frac{1}{2}
   \sum_{\mu\nu}
   \tilde{D}^{\text{2c}}_{\nu\mu}
   \underbrace{
   \sum_{\kappa\lambda}
   G ^{\text{2c}}_{\mu\nu,\kappa\lambda}
   \tilde{D}^{\text{2c}}_{\lambda\kappa}
   }_{F^{\text{2c,2e}}_{\mu\nu}}
   +
   \frac{1}{2}
   \sum_{\mu\nu}
   \tilde{D}^{\text{2c}}_{\nu\mu}
   \underbrace{
   \sum_{\kappa\lambda}
   \Delta\tilde{G}^{\text{2c}}_{\mu\nu,\kappa\lambda}
   \tilde{D}^{\text{2c}}_{\lambda\kappa}
   }_{\Delta\tilde{F}^{\text{2c,2e}}_{\mu\nu}}
   .
\end{align}

We now seek a suitable approximation for the differential two-electron Fock matrix $\Delta\tilde{\vec{F}}^{\text{2c,2e}}$. In line with previous authors we exploit the expected local atomic nature
of the two-electron picture-change corrections, but we will impose the condition that the scheme should reproduce \textit{atomic} 4c SCF calculations \textit{exactly} at the 2c level. We accordingly start from a superposition of
converged atomic quantities rather than the converged molecular one, \emph{i.e.}
\begin{align} \label{eq:amfDeltaF2}
  \Delta\tilde{\vec{F}}^{\text{2c,2e}}
  \simeq
  \Delta\tilde{\vec{F}}^{\text{2c,2e}}_{\bigoplus}
  =
  \bigoplus_{K=1}^{M}
  \Delta\tilde{\vec{F}}^{\text{2c}}_{K}[\tilde{\vec{D}}_K^{\text{2c}}],
\end{align}
where $K$ runs over all atoms in an $M$-atomic system.
Such an approach defines our {\it{atomic mean-field exact two-component}} scheme, denoted as amfX2C. Due to the atomic nature of 
amfX2C two-electron picture-change corrections, their evaluation scales linearly with the system size 
(or sub-linearly if there are multiple instances of an atomic type).
To summarize the essentials, we propose the following computational scheme to arrive at the amfX2C model:
\begin{enumerate}
\item For each atomic type $K$ we perform a 4c Kramers-restricted (KR) average-of-configuration (AOC) HF calculation~\cite{Thyssen_phd2004}
-- or, if the latter is not available, -- a 4c KR fractional occupation HF calculation.
\item The converged atomic Fock matrix $\vec{F}_K^{\text{4c}}$ is exactly block-diagonalized to give its 2c counterpart $\tilde{\vec{F}}_K^{\text{2c}}$\ as well as picture-changed coefficients $\tilde{\vec{c}}_K^{\text{2c}}$ 
and density matrix $\tilde{\vec{D}}_K^{\text{2c}}$.
\item Using the latter quantity, we build the atomic 2c Fock matrix $\vec{F}^{\text{2c}}_{K}[\tilde{\vec{D}}_K^{\text{2c}}]$ 
      with untransformed two-electron integrals, Eq.~\eqref{eq:2cFock_nopoc}.
\item The differential atomic Fock matrix $\Delta\tilde{\vec{F}}^{\text{2c}}_K[\tilde{\vec{D}}_K^{\text{2c}}]$ 
      is now built according to Eq.~\eqref{eq:DeltaF2}.
\item The atomic matrices $\Delta\tilde{\vec{F}}^{\text{2c}}_{K}$ and $\vec{F}^{\text{4c,2e}}_{K}$ are then inserted
      in the appropriate atomic blocks to form approximate molecular two-electron picture-change correction matrix ($\Delta\tilde{\vec{F}}^{\text{2c,2e}}_{\bigoplus}$), Eq.~\eqref{eq:amfDeltaF2}, and 
      approximate molecular two-electron Fock matrix ($\vec{F}^{\text{4c,2e}}_{\bigoplus}$), respectively:
\begin{align} \label{}
  \vec{F}^{\text{4c,2e}}
  \simeq
  \vec{F}^{\text{4c,2e}}_{\bigoplus}
  =
  \bigoplus_{K=1}^{\text{M}}
  \vec{F}^{\text{4c,2e}}_{K}[\vec{D}_K^{\text{4c}}].
\end{align}
\item The molecular X2C decoupling matrix $\vec{U}$ is built from $\vec{h}^{\text{4c}} + \vec{F}^{\text{4c,2e}}_{\bigoplus}$.  
\item Finally, SCF iterations are carried out with amfX2C expressions that approximate the exact molecular Fock matrix and 
energy expressions
\begin{align} \label{}
   \tilde{F}^{\text{2c}}_{\mu\nu}
   &\simeq
   \tilde{F}^{\text{amfX2C}}_{\mu\nu}
   =
   \underbrace{
   \tilde{h}^{\text{2c}}_{\mu\nu}
   +
   \Delta\tilde{F}^{\text{2c,2e}}_{\bigoplus,\mu\nu}
   }_{\text{static term}}
   +
   \underbrace{
   F^{\text{2c,2e}}_{\mu\nu}[\tilde{\vec{D}}^{\text{2c}}]
   }_{\text{dynamic term}}
   \\
   \tilde{E}^{\text{2c}}
   &\simeq
   \tilde{E}^{\text{amfX2C}}
   = 
   \sum_{\mu\nu}
   \tilde{D}^{\text{2c}}_{\nu\mu}
   \Big(
   \tilde{h}^{\text{2c}}_{\mu\nu}
   +
   \frac{1}{2}\Delta\tilde{F}^{\text{2c,2e}}_{\bigoplus,\mu\nu}
   +
   \frac{1}{2}
   F^{\text{2c,2e}}_{\mu\nu}[\tilde{\vec{D}}^{\text{2c}}]
   \Big).
\end{align}
\end{enumerate}
A pseudo-code describing the essential steps of our amfX2C approach for both HF and 
Kohn--Sham DFT theory is listed in Alg.~\ref{alg:amfX2C}.

\subsection{The amfX2C Hamiltonian -- Kohn--Sham DFT framework}\label{sec:theory:KS}

Section \ref{sec:theory:HF} has so far exclusively focused on a discussion of 2ePCE corrections 
within a mean-field HF scheme. As indicated in Algorithm \ref{alg:amfX2C}, the proposed amfX2C scheme has also the appealing feature that it can straightforwardly be extended to a KS-DFT framework.

\subsubsection{The closed-shell case}

Let us start for simplicity by considering the  closed-shell molecular case  where  
the 4c energy and KS matrix read
\begin{align} \label{eq:4cFockKS}
  E^{\text{4c}}
  &=
  \sum_{\mu\nu}h^{\text{4c}}_{\mu\nu}D^{\text{4c}}_{\nu\mu}
  +
  \frac{1}{2}\sum_{\mu\nu\kappa\lambda}D^{\text{4c}}_{\nu\mu}G^{\omega;\text{4c}}_{\mu\nu,\kappa\lambda}D^{\text{4c}}_{\lambda\kappa}
  +
  E_{xc}^{\text{4c}}\left[n^{\text{4c}}\right]
  \\
  F^{\text{4c}}_{\mu\nu}
  &=
  h^{\text{4c}}_{\mu\nu}
  +
  \sum_{\kappa\lambda}G^{\omega;\text{4c}}_{\mu\nu,\kappa\lambda}D^{\text{4c}}_{\lambda\kappa}
  +
  \int v_{xc}\left[n^{\text{4c}}\right]\left(\boldsymbol{r}\right)\Omega_{\mu\nu}^{\text{4c}}\left(\boldsymbol{r}\right)d^{3}\boldsymbol{r}
  ;\quad 
  v_{xc}\left[n\right]\left(\boldsymbol{r}\right)=\frac{\delta E_{xc}}{\delta n\left(\boldsymbol{r}\right)}.
\end{align}  
Here, we have generalized the anti-symmetrized two-electron AO integrals of Eq.~\eqref{eq:eri} to include the weight $\omega$ of exact exchange. As usual, $E_{xc}$ and $v_{xc}$ refer to the exchange--correlation energy functional and the corresponding potential, respectively. 

Formally $E_{xc}$ may be expressed as an integral over an xc energy density $\varepsilon_{xc}$
\begin{equation}
    E_{xc}[n]=\int\varepsilon_{xc}\left[n\right]\left(\boldsymbol{r}\right)d^{3}\boldsymbol{r},
\end{equation}
which is itself a functional of the number density ($n$). This allows for instance the electron number to be known locally such that the derivative discontinuity can be obeyed.\cite{Perdew_PRL1982}  Crucial for the following, though, is that density functional approximations (DFA) employ \textit{local} \textit{ans\"atze}. For instance, on the second rung of the “Jacob's ladder” of DFA~\cite{perdew:ladder,Perdew_JCP2005} we find the generalized gradient approximation (GGA)
\begin{equation}\label{eq:GGA}
    E_{xc}[n]
    =
    \int\varepsilon_{xc}^{\text{GGA}}
    \left(n\left(\boldsymbol{r}\right),g\left(\boldsymbol{r}\right)\right)
    d^{3}\boldsymbol{r};\quad g\left(\boldsymbol{r}\right)=\nabla n\left(\boldsymbol{r}\right)\cdot\nabla n\left(\boldsymbol{r}\right),
\end{equation}
where each integration point just needs local input.

Proceeding at the GGA/hybrid level, 
we find that the picture-changed KS matrix can be expressed as
\begin{equation} 
   \tilde{F}^{\text{2c}}_{\mu\nu}  
   = 
   \tilde{h}^{\text{2c}}_{\mu\nu} 
   +
   \sum_{\kappa\lambda} \tilde{G}^{\omega;\text{2c}}_{\mu\nu,\kappa\lambda} \tilde{D}^{\text{2c}}_{\lambda\kappa}
   + 
   \int v_{xc}^{\text{GGA}}\left(n^{4c}\left(\boldsymbol{r}\right),g^{4c}\left(\boldsymbol{r}\right)\right)\tilde{\Omega}_{\mu\nu}^{\text{2c}}\left(\boldsymbol{r}\right)d^{3}\boldsymbol{r}.
\end{equation}
We again recover an expression in terms of picture-changed quantities, but the xc potential is seen to still use 4c variables as input. However, proceeding as in the HF case (c.f.~Eq.(\ref{eq:HF-4c-2c-energy})\ in Section \ref{sec:theory:HF}), these can be re-expressed in terms of 2c quantities
\begin{align}
  n^{\text{4c}}\left(\boldsymbol{r}\right) 
  &= 
  \sum_{XY}\sum_{\mu\nu}\left[\Omega^{\text{4c}}\left(\boldsymbol{r}\right)\right]_{\mu\nu}^{\text{XY}}\left[D^{\text{4c}}\right]_{\nu\mu}^{\text{YX}}
  =
  \sum_{\mu\nu}\left[\tilde{\Omega}^{\text{2c}}\left(\boldsymbol{r}\right)\right]_{\mu\nu}\left[\tilde{D}^{\text{2c}}\right]_{\nu\mu}
  =
  \tilde{n}^{\text{2c}}\left(\boldsymbol{r}\right),
  \\[0.2cm]
  g^{\text{4c}}\left(\boldsymbol{r}\right) 
  &= 
  \nabla\tilde{n}^{\text{2c}}\left(\boldsymbol{r}\right)\cdot\nabla\tilde{n}^{\text{2c}}\left(\boldsymbol{r}\right)
  =
  \tilde{g}^{\text{2c}}\left(\boldsymbol{r}\right).
\end{align}
This also means that the xc energy and potential can be expressed entirely in terms of 2c quantities
\begin{align}
    \tilde{E}_{xc}^{\text{2c}}
    & =
    \int\varepsilon_{xc}^{\text{GGA}}\left(\tilde{n}^{\text{2c}}\left(\boldsymbol{r}\right),\tilde{g}^{\text{2c}}\left(\boldsymbol{r}\right)\right)d^{3}\boldsymbol{r},
    \\
    \tilde{F}^{\text{2c,xc}}_{\mu\nu}
    & =
    \int v_{xc}^{\text{GGA}}\left(\tilde{n}^{2c}\left(\boldsymbol{r}\right),\tilde{g}^{2c}\left(\boldsymbol{r}\right)\right)\tilde{\Omega}_{\mu\nu}^{\text{2c}}\left(\boldsymbol{r}\right)d^{3}\boldsymbol{r}.
\end{align}
In passing we note that the direct use of the GGA xc potential leads to contributions on the form
\begin{equation}\label{eq:GGA1}
   F_{\mu\nu}^{\text{xc}}
   =
   \int v_{xc}^{\text{GGA}}\left(\boldsymbol{r}\right)\Omega_{\mu\nu}\left(\boldsymbol{r}\right)d^{3}\boldsymbol{r}
   ;\quad 
   v_{xc}^{\text{GGA}}\left(\boldsymbol{r}\right)
   =
   \left[\frac{\partial\varepsilon_{xc}^{\text{GGA}}}{\partial n}
   -
   2\boldsymbol{\nabla}\cdot\left(\left(\frac{\partial\varepsilon_{xc}^{\text{GGA}}}{\partial g}\right)\boldsymbol{\nabla} n\right)\right]\left(\boldsymbol{r}\right).
\end{equation}
However, the second term of the GGA potential will require the expensive calculation of the Hessian $\nabla^{2}n$ of the number density, so
usually a derivative is shifted over to the overlap distribution $\Omega_{\mu\nu}\left(\boldsymbol{r}\right)$, using integration by parts, giving
\begin{equation}\label{eq:GGA2}
  F_{\mu\nu}^{\text{xc}}
  =
  \int
  \left[
  \frac{\partial\varepsilon_{xc}^{\text{GGA}}}{\partial n}\Omega_{\mu\nu}\left(\boldsymbol{r}\right)
  +
  2\left(\frac{\partial\varepsilon_{xc}^{\text{GGA}}}{\partial g}\right)(\boldsymbol{\nabla}n)\cdot\boldsymbol{\nabla}\Omega_{\mu\nu}\left(\boldsymbol{r}\right)
  \right]
  d^{3}\boldsymbol{r}.
\end{equation}
These manipulations commute with the picture-change transformation, though, and so we will for simplicity  continue with the form of Eq.~\eqref{eq:GGA1}.

Just as in the case of HF we will argue that, if the 2c calculation is carried out with the correctly transformed overlap distribution $\tilde{\Omega}_{\mu\nu}^{2c}\left(\boldsymbol{r}\right)$,
in addition to the picture-changed one- and two-electron integrals, it will converge to the picture-changed coefficients $\tilde{c}^{\text{2c}}$ obtained from the corresponding 4c calculation. However, again the correct decoupling matrix $\vec{U}$, namely the one associated with the \textit{converged} KS matrix, is not available at the start of calculations and so we will have to seek approximations. One option, pursued by Iakabata and Nakai,\cite{Ikabata_PCCP2021} is to use the decoupling matrix $\vec{U}$\ associated with the Dirac Hamiltonian instead. The point-wise picture-change transformation of the overlap distribution, even with local approximations, adds significant computational cost, though, and the chosen decoupling matrix $\vec{U}$\ is not optimal. An alternative would be to make picture-change corrections to the number density, starting from
\begin{equation}
    \Delta\tilde{n}^{\text{2c}}\left(\boldsymbol{r}\right)
    =
    \tilde{n}^{\text{2c}}\left(\boldsymbol{r}\right)-n^{\text{2c}}\left(\boldsymbol{r}\right)
    =
    n^{\text{4c}}\left(\boldsymbol{r}\right)-n^{\text{2c}}\left(\boldsymbol{r}\right).
\end{equation}
Due to the local nature of the corrections we would expect these  corrections to be separable into atomic contributions, possibly approximated by model densities (see e.g. Refs.~\citenum{vanW05a,peng07b}), that is
\begin{equation}
    \Delta\tilde{n}^{\text{2c}}\left(\boldsymbol{r}\right)
    \simeq
    \sum_{K=1}^{M}\Delta\tilde{n}_{K}^{\text{2c}}\left(\boldsymbol{r}\right).
\end{equation}
Here it is important to stress that the atomic number density ${n}_K^{\text{2c}}$ (without the tilde) is untransformed in the sense that it employs an \textit{untransformed} overlap distribution matrix $\vec{\Omega}^{\text{2c}}_{K}$, but the \textit{correctly transformed} coefficients 
$\big\{\tilde{\vec{c}}^{\text{2c}}_{K,i}\big\}$ corresponding to the parent 4c atomic calculation. 
Since we expect $\Delta\tilde{n}_{K}^{\text{2c}}\left(\boldsymbol{r}\right)$
to be non-zero only in the deep atomic core, one could exploit spherical symmetry by calculating the correction on a \textit{radial} grid. However, we have not pursued this approach, since it still involves point-wise corrections, albeit over a significantly reduced number of integration points.

Instead, we propose the following scheme which integrates nicely with the scheme proposed for HF: for 
each atomic species $K$ we run a 4c KR fractional occupation KS-calculation which provides the converged atomic KS matrix $\vec{F}_{K}^{\text{4c}}$. From it we can 
directly extract the atomic decoupling matrix $\vec{U}_K$\ and the corresponding picture-changed KS-matrix $\tilde{\vec{F}}_{K}^{\text{2c}}$, notably containing $\tilde{\vec{F}}_{K}^{\text{2c,xc}}$. We next build the untransformed equivalent
\begin{equation}
  F_{K;\mu\nu}^{\text{2c,xc}}
  =
  \int v_{xc}^{\text{GGA}}\left(n_{K}^{\text{2c}}\left(\boldsymbol{r}\right),g_{K}^{\text{2c}}\left(\boldsymbol{r}\right)\right)
  \Omega_{K;\mu\nu}^{\text{2c}}\left(\boldsymbol{r}\right)d^{3}\boldsymbol{r},
\end{equation}
using the correctly picture-changed transformed coefficients $\tilde{\vec{c}}_{K}^{\text{2c}}$. Our amfX2C picture-change correction 
to the xc potential is then obtained from atomic quantities as
\begin{equation}
  \Delta\tilde{\vec{F}}^{\text{2c,xc}}
  \simeq
  \Delta\tilde{\vec{F}}^{\text{2c,xc}}_{\bigoplus}
  =
  \bigoplus_{K=1}^{M}\Delta\tilde{\vec{F}}_{K}^{\text{2c,xc}}
  ;\qquad
  \Delta\tilde{\vec{F}}_{K}^{\text{2c,xc}}=\tilde{\vec{F}}_{K}^{\text{2c,xc}}-\vec{F}_{K}^{\text{2c,xc}}.
\end{equation}
Similarly, the xc energy is corrected by first writing 
$\tilde{E}_{xc}^{\text{2c}} = E_{xc}^{\text{2c}} + \Delta\tilde{E}_{xc}^{\text{2c}}$,
and then seeking an atomic approximation to the correction
\begin{equation}
  \Delta\tilde{E}_{xc}^{\text{2c}}
  =
  \int\varepsilon_{xc}^{\text{GGA}}
  \left(\tilde{n}^{\text{2c}}\left(\boldsymbol{r}\right),\tilde{g}^{\text{2c}}\left(\boldsymbol{r}\right)\right)
  d^{3}\boldsymbol{r}
  -
  \int\varepsilon_{xc}^{\text{GGA}}
  \left(n^{\text{2c}}\left(\boldsymbol{r}\right),g^{\text{2c}}\left(\boldsymbol{r}\right)\right)
  d^{3}\boldsymbol{r}.
\end{equation}
This results in our amfX2C picture-change correction 
to the xc energy
\begin{equation}
  \Delta\tilde{E}_{xc}^{\text{2c}}
  \simeq
  \Delta\tilde{E}_{xc,\bigoplus}^{\text{2c}}
  =
  \sum_{K=1}^{M}\left(\tilde{E}_{xc;K}^{\text{2c}} - E_{xc;K}^{\text{2c}}\right).
\end{equation}
At first sight this looks like a rather poor approximation, since, clearly
\begin{equation}
 \sum_{K}\tilde{E}_{xc;K}^{\text{2c}}
 =
 \sum_{K}\int\varepsilon_{xc}^{\text{GGA}}
 \left(\tilde{n}_{K}^{\text{2c}}\left(\boldsymbol{r}\right),\tilde{g}_{K}^{\text{2c}}\left(\boldsymbol{r}\right)\right)
 d^{3}\boldsymbol{r}
 \ne
 \int\varepsilon_{xc}^{\text{GGA}}
 \Big(\sum_{K}\tilde{n}_{K}^{\text{2c}}\left(\boldsymbol{r}\right),\sum_{K}\tilde{g}_{K}^{\text{2c}}\left(\boldsymbol{r}\right)\Big)
  d^{3}\boldsymbol{r},
\end{equation}
due to the general non-linear form of the xc functionals. However,
we are calculating picture-change \textit{corrections}, and so one may expect that points for which $\tilde{n}_{K}^{2c}\left(\boldsymbol{r}\right)-n_{K}^{2c}\left(\boldsymbol{r}\right)$
deviates significantly from zero for some atomic species $K$ does not overlap with equivalent points for any other species. Under such conditions our approximation becomes perfectly valid due to the local
ansatz of the energy density $\varepsilon_{xc}$, cf. Eq.~\eqref{eq:GGA}.

\subsubsection{The noncollinear open-shell case}
So far, we have discussed the KS amfX2C approach for a closed-shell molecular system which is characterized 
by a time--reversal symmetric density matrix. Due to the symmetry, the entire dependence of 
the exchange--correlation energy density reduces for a local-density approximation (LDA) only to the number density 
($n$) [see Eq.~\eqref{eq:4cFockKS}], for a generalized-gradient approximation (GGA) also to its gradient, 
$g_{nn}\equiv(\vect\nabla n)\!\cdot\!(\vect\nabla n)$. 

The situation is more complex for open-shell systems, where a general Kramers-unrestricted 
formalism results in a density matrix that has both the time-reversal symmetric (TRS)
as well as time-reversal antisymmetric (TRA) component.~\cite{komo19a,ReSpect2020} In fact, the latter component gives rise to a non-zero 
electron spin density, whose z-component ($s_{z}$) enters together with its gradient ($\vect\nabla{s_{z}}$) into the non-relativistic 
exchange--correlation energy expression, i.e.
$\varepsilon_{xc}^{\text{GGA}} \equiv \varepsilon_{xc}^{\text{GGA}} \big( \left\{\rho(\vec{r}\right\}\big), \rho=n, g_{nn}, s_{z}, (\vect\nabla{s_{z}}\!\cdot\!\vect\nabla{s_{z}}), (\vect\nabla{n}\!\cdot\!\vect\nabla{s_{z}})$.

However, the presented parametrization of the exchange--correlation energy involving  
only the $z$-component of the electron spin density and its gradient 
is inadequate for theories including the spin--orbit interaction, 
since the spatial and spin degrees of freedom are no longer independent. 
Their coupling results in a lack of rotational invariance of the exchange--correlation energy
if only $z$ spin-components are involved. This variance can be circumvented by 
a \emph{noncollinear} parametrization/generalization of the non-relativistic exchange--correlation energy density.

A common noncollinear ansatz follows earlier LDA-based works of Kubler \emph{et al.},~\cite{Kubler1988} 
Sandratskii,~\cite{Sandratskii1998} and van Wuellen~\cite{VanWuellen2002}
where the variable $s_z$ is replaced by its corresponding magnitude $|\vect{s}|$. 
Although this extension possesses no numerical problems in the 
evaluation of exchange--correlation energy, noncollinear potentials and kernels derived from GGA-type functionals 
are prone to numerical instabilities.~\cite{komo19a} A more recent approach, which has been adopted in this work, is based on the noncollinear ansatz proposed by Scalmani and Frisch,~\cite{Scalmani2012} where variables depending on the 
$z$ quantization axis are substituted by more adequate rotationally invariant counterparts:
\begin{align}
\label{eq:noncoll-ansatz-2}
\begin{split}
\begin{array}[c]{ccc}
  s_z \rightarrow s\equiv|\vect{s}|;
  \quad
  (\vect\nabla{s_z})\!\cdot\!(\vect\nabla{s_z})
  \rightarrow
  g_{ss} \equiv \displaystyle\sum_{k}(\vect\nabla{s_k})\cdot(\vect\nabla{s_k});
  \quad
  (\vect\nabla{n})\!\cdot\!(\vect\nabla{s_z}) 
  \rightarrow 
  g_{ns} \equiv f_{\nabla} g.
\end{array}
\end{split}
\end{align}
Here, $k\in x,y,z$; $g\equiv|\vect g|$ with $g_k=(\vect\nabla{n})\!\cdot\!(\vect\nabla{s_k})$, 
and $f_{\nabla}=\mathrm{sgn}(\vect g \cdot \vect{s})$.
The noncollinear exchange--correlation energy then reads
\begin{align}
\label{eq:os-Exc-S}
\begin{split}
   E_{xc} 
   = 
   \int \varepsilon_{xc}^{\text{GGA}} \big(\left\{\rho(\vec{r}\right\} \big) \dV{r},\quad\rho=n, g_{nn}, s, g_{ss}, g_{ns} \,
\end{split}
\end{align}
whereas the noncollinear exchange--correlation potential has the form~\cite{komo19a}
\begin{align}
\begin{split}
 \label{eq:os-Vxc-S}
   F^{\text{xc}}_{\mu\nu} 
   =
   \frac{\mathrm{d}E_{xc}}{\mathrm{d}D_{\nu\mu}} 
   & = \int \bigg(
       v_{xc}^n                                \,             \Omega^0_{\mu\nu}
     + v_{xc}^s        \sum_{k}\frac{s_k}{s}   \,             \Omega^k_{\mu\nu}
     +2v_{xc}^{g_{nn}} \sum_{k}(\nabla_{\!k}n) \,\nabla_{\!k} \Omega^0_{\mu\nu}
   \\     
   & +
    2v_{xc}^{g_{ss}} \sum_{k,l}(\nabla_{\!l}s_k) \,\nabla_{\!l} \Omega^k_{\mu\nu}
   + v_{xc}^{g_{ns}} \sum_{k,l}f_\nabla \,\frac{g_k}{g} \left[ \,(\nabla_{\!l}s_k) \,\nabla_{\!l} \Omega^0_{\mu\nu} +
                                                               \,(\nabla_{\!l}n)   \,\nabla_{\!l} \Omega^k_{\mu\nu}
                                                               \right]
         \bigg)
   \dV{r}.
\end{split}
\end{align}
Here, $k,l\in x,y,z$ and $v_{xc}^{\rho}$ refer to the partial derivative of $\varepsilon_{xc}^{\text{GGA}}$ with respect to
$\rho\in n, g_{nn}, s, g_{ss}, g_{ns}$. $\Omega^{0}_{\mu\nu}$ and $\Omega^{k}_{\mu\nu}$ stand for the 
overlap and spin distribution functions, respectively, the latter being defined similarly to
$\Omega^{0}_{\mu\nu}$ in Eq.~\eqref{eq:omega0} as 
\begin{equation} \label{eq:omegak}
   \Omega_{\mu\nu}^{k}(\vec{r})
   \equiv 
   \chi_{\mu}^{\dagger}(\vec{r})\hat{\Sigma}_{k}\chi_{\nu}(\vec{r})
\end{equation}
and involves components of the electron spin operator $\hat{\vec{\Sigma}}$.~\cite{komo19a}
Note that the evaluation of the exchange--correlation potential in Eq.~\eqref{eq:os-Vxc-S} requires special attention to 
the limiting cases when the $s$ or $g$ functions approach zero. A detailed description of such a procedure is 
given in Ref.~\citenum{komo19a}.

\begin{figure}
\begin{algorithm}[H]
\caption{Pseudo-code highlighting the essential steps for the amfX2C approach.}
\label{alg:amfX2C}
\begin{algorithmic}[1]
  \State \mycm{Initialize the \textcolor{blue}{molecular} two-electron (2e) Fock matrices and XC energy}
  \State $
         \vec{F}^{\text{4c,2e}}_{\bigoplus}=\vec{0};
         \quad\
         \Delta\tilde{\vec{F}}^{\text{2c,2e}}_{\bigoplus}=\vec{0}; 
         \quad\ 
         \Delta\tilde{E}_{xc,\bigoplus}^{2c}=0
         $
  \ForAll{\texttt{unique atom types $K$\ $\in$ molecule}}
    \State \textsf{Let} $\{\mu, \nu\} \in {\rm atomic\ basis}\ K$
    \State \mycm{Solve the 4c SCF equation}
    \State          $
               \vec{F}^{\text{4c}}_{K}\vec{c}^{\text{4c}}_{K}
                =  
               \vec{c}^{\text{4c}}_{K}\epsilon^{\text{4c}}_{K}$
    \,\, \textsf{with}
    \,\,
            $ F^{\text{4c}}_{K,\mu\nu}
              = 
              \begin{cases}
                  F^{\text{4c,HF}}_{K,\mu\nu}[\vec{D}^{\text{4c}}_{K}] 
                  = 
                  h^{\text{4c}}_{K,\mu\nu}
                  +
                  \sum\limits_{\gamma, \delta \in K}
                  G_{\mu\nu,\gamma\delta}^{\text{4c}}D^{\text{4c}}_{K,\delta\gamma}
                  \\[0.2cm]
                  F^{\text{4c,KS}}_{K,\mu\nu}[\vec{D}^{\text{4c}}_{K}] 
                  = 
                  h^{\text{4c}}_{K,\mu\nu}
                  +
                  \sum\limits_{\gamma, \delta \in K}
                  G^{\omega,\text{4c}}_{\mu\nu,\gamma\delta}D^{\text{4c}}_{K,\delta\gamma}
                  +
                  F^{\text{4c,xc}}_{K,\mu\nu}[\vec{D}^{\text{4c}}_{K}]
              \end{cases}
             $
             
        \State \mycm{Add $K$-th \textcolor{blue}{atomic} 2e Fock contrib. $\vec{F}^{\text{4c,2e}}_{K}$ to the corresponding molecular block}
        \State \label{alg:amfX2c:4cFockblocked} $
               \vec{F}^{\text{4c,2e}}_{\bigoplus}
               \gets
               \vec{F}^{\text{4c,2e}}_{K}
               $
        \,\, \textsf{with}
        \,\, $
               F^{\text{4c,2e}}_{K,\mu\nu} =  
               \begin{cases}  F^{\text{4c,HF}}_{K,\mu\nu}[\vec{D}^{\text{4c}_{K}}] - h^{\text{4c}}_{K,\mu\nu}
                     \\[0.2cm]
               F^{\text{4c,KS}}_{K,\mu\nu}[\vec{D}^{\text{4c}}_{K}] - h^{\text{4c}}_{K,\mu\nu}
               \end{cases}
               $
        \State \mycm{Evaluate the \textcolor{blue}{atomic} X2C decoupling matrix $\vec{U}_{K}$ from  $\vec{F}^{\text{4c}}_{K}$ and calculate}
        \State
        $
        \tilde{\vec{D}}^{\text{2c}}_{K}
               =
               \left[ \vec{U}^{\dagger}_{K} \vec{D}^{\text{4c}}_{K} \vec{U}_{K} \right]^{\text{LL}}
               ;\qquad 
               \Delta\tilde{\vec{F}}^{\text{2c,2e}}_{K}
               =
               \left[ \vec{U}^{\dagger}_{K} \vec{F}^{\text{4c,2e}}_{K} \vec{U}_{K} \right]^{\text{LL}}
               -
               \vec{F}^{\text{2c,2e}}_{K}
        $
        \State \mycm{where the latter term facilitates untransformed quantities $G^{\text{2c}}$, $G^{\omega,\text{2c}}$, and $F^{\text{2c,xc}}_{K}$}
        \State
        $
        F^{\text{2c,2e}}_{K,\mu\nu}
               =
               \begin{cases} F^{\text{2c,2e,HF}}_{K,\mu\nu}[\tilde{\vec{D}}^{\text{2c}}_{K}] 
                     = 
                     \sum\limits_{\gamma, \delta \in K}
                     G_{\mu\nu,\gamma\delta}^{\text{2c}}\tilde{D}^{\text{2c}}_{K,\delta\gamma}
                     \\[0.2cm]
                     F^{\text{2c,2e,KS}}_{K,\mu\nu}[\tilde{\vec{D}}^{\text{2c}}_{K}] 
                     = 
                     \sum\limits_{\gamma, \delta \in K}
                     G^{\omega,\text{2c}}_{\mu\nu,\gamma\delta}\tilde{D}^{\text{2c}}_{K,\delta\gamma}
                     +
                     F^{\text{2c,xc}}_{K,\mu\nu}[\tilde{\vec{D}}^{\text{2c}}_{K}]
               \end{cases}
        $
       \State \mycm{Add $K$-th \textcolor{blue}{atomic} block of the picture-change error correction to the corresponding \textcolor{blue}{molecular} block. In case of DFT, add also the \textcolor{blue}{atomic} XC energy correction:}
       \State \label{alg:amfx2c:xcenergy}
        $
        \Delta\tilde{\vec{F}}^{\text{2c,2e}}_{\bigoplus} \leftarrow \Delta\tilde{\vec{F}}^{\text{2c,2e}}_{K}
        ;\qquad
        \Delta\tilde{E}_{xc,\bigoplus}^{\text{2c}} 
        \gets \left(E_{xc,K}^{\text{4c}}[\vec{D}_{K}^{\text{4c}}]-E_{xc,K}^{\text{2c}}[\tilde{\vec{D}}_{K}^{\text{2c}}]\right)
        $
    \EndFor
    \State \textsf{Let} $\{\mu, \nu\} \in \text{full molecular basis}$
    \State \mycm{Evaluate the \textcolor{blue}{molecular} X2C decoupling matrix $\vec{U}$ from}
   \State
   $
      \tilde{\vec{h}}^{\text{4c}} 
      = 
      \vec{h}^{\text{4c}} 
      + 
      \vec{F}^{\text{4c,2e}}_{\bigoplus}
   $
    \State \mycm{Solve the 2c SCF equation with the amfX2C Fock matrix operator}
    \State
    $
      \vec{F}^{\text{2c}}\vec{c}^{\text{2c}}
      =
      \vec{c}^{\text{2c}}\epsilon^{\text{2c}}
      \,\, \textsf{with} \,\,
      F^{\text{2c}}_{\mu\nu}
      \equiv
      F^{\text{amfX2C}}_{\mu\nu}
      =
      \underbrace{\left[ \vec{U}^{\dagger} \vec{h}^{\text{4c}} \vec{U} \right]^{\text{LL}}_{\mu\nu}
      +
      \Delta\tilde{F}^{\text{2c,2e}}_{\bigoplus,\mu\nu}}_{\rm static\ terms}
      +
      \underbrace{F^{\text{2c,2e}}_{\mu\nu}[\vec{D}^{2c}]}_{\rm dynamic\ term}
   $
  \end{algorithmic}
\end{algorithm}
\end{figure}

\subsection{Extended amfX2C Hamiltonian}\label{sec:eamf-and-remarks}

Having introduced the amfX2C scheme for both HF and KS mean-field theories, let us conclude this theory section by commenting on some important aspects of the amfX2C scheme, as well as comparing it to existing models for 2ePCE corrections. Ultimately, the discussion leads to 
the introduction of an extended amfX2C model, dubbed eamfX2C, which has the potential to outperform the amfX2C model, for instance, in properly treating long-range Coulomb interactions in solids.

We start by noting that: 
(i) in contrast to Liu and Cheng~\cite{liuj18} our amfX2C scheme allows to take into account PCE corrections for 
both \textit{spin-independent} and \textit{spin-dependent} parts of the two-electron interaction; 
(ii) the proposed amfX2C approach has the additional appealing feature that it allows its straightforward extension to a KS-DFT framework as discussed in Section~\ref{sec:theory:KS};
(iii) the algebraic nature of amfX2C also allows an easy extraction of 2ePCE corrections not only from the common 
2e Coulomb interaction term but also from more elaborate Gaunt and Breit 2e-interaction terms;
(iv) the 2ePCE corrections are only introduced in the atomic diagonal blocks. This further implies:
\begin{itemize}
   \item The 2ePCE corrections will not contribute to the molecular gradient.
   \item The direct 2e Coulomb contribution will not cancel exactly the electron-nucleus interaction at long distance from atomic centers that potentially prevents a direct application of amfX2C in solid-state calculations. This issue was discussed for instance by van W{\"u}llen and Michauk, and solved by building the former contributions using a superposition of atomic model densities~\cite{vanW05a}, 
   although such a scheme does not accommodate HF exchange contributions. 
 \end{itemize}
   In order to overcome the latter, particular shortcoming of the amfX2C model, we additionally propose a modified amfX2C model which exploits a superposition of atomic density matrices. 
   The resulting 
   \emph{extended} amfX2C model (eamfX2C) is summarized in Alg.~\ref{alg:eamfX2C}. 
   Most importantly, in contrast to the amfX2C model, where we assemble a molecular 4c Fock matrix $\vec{F}^{\text{4c}}_{\bigoplus}$\ from atomic building blocks (see line \ref{alg:amfX2c:4cFockblocked} in Alg.~\ref{alg:amfX2C}), this task is replaced in the eamfX2C algorithm by the buildup of a molecular \textit{density} matrix $\vec{D}^{\text{4c}}_{\bigoplus}$\ from atomic density matrices as indicated in line~\ref{alg:eamfX2c:4cDensblocked} of Alg.~\ref{alg:eamfX2C}. The latter construction therefore entails the evaluation of a two-electron (KS-)Fock matrix contribution in the \textit{full} molecular basis within a 4c framework (c.f.~line \ref{alg:eamfX2C:line4cF} of Alg~.\ref{alg:eamfX2C}) which is absent in the molecular computational panel (lower part of Alg.~\ref{alg:amfX2C}) of the simpler amfX2C model. Although introducing such a requirement seems odd at a first glance, in particular, with regard to the computational scaling, let us recall that an efficient density-based screening in the two-electron (KS-)Fock matrix construction will enable a calculation of the term $F_{\mu\nu}^{\text{4c,2e}}[\vec{D}^{\text{4c}}_{\bigoplus}]$\ at a fractional cost of a regular two-electron (KS-)Fock matrix evaluation because of the sparsity associated with the molecular density matrix $\vec{D}^{\text{4c}}_{\bigoplus}$. 
   In this regard, one can recognize a similarity between the eamfX2C scheme and the atomic initial guess proposed by van Lenthe and co-workers~\cite{vanLenthe2006} where the initial Fock matrix is formed from a superposition of atomic density matrices.
   Moreover, in the KS-DFT framework, one can also easily obtain the xc energy picture-change correction (Alg.~\ref{alg:eamfX2C}, line \ref{alg:eamfX2c:xcenergy}) from contributions evaluated in the full \textit{molecular}\ basis, 
   \begin{equation}
     \Delta\tilde{E}^{\text{2c}}_{xc} 
     \simeq 
     \tilde{E}^{\text{2c}}_{xc,\bigoplus} 
     =
     {E}^{\text{4c}}_{xc}[{\vec{D}_{\bigoplus}^{\text{4c}}}] - {E}^{\text{2c}}_{xc}[{\tilde{\vec{D}}_{\bigoplus}^{\text{2c}}}]\ ,
   \end{equation}
   in contrast to the correction term $\Delta \tilde{E}^{\text{2c}}_{xc}$ of the amfX2C model (Alg.~\ref{alg:amfX2C}, line \ref{alg:amfx2c:xcenergy}) which consists of a sum of $K$\  contributions each calculated in the $K$-th \textit{atomic} basis.

\subsection{A remark on notations}\label{sec:notations}

Since the combination of a several 2ePCE correction models with a multiple defining Hamiltonians 
for obtaining the unitary decoupling matrix $\vec{U}$\ may easily lead to confusion,
we have decided to introduce a notation for X2C Hamiltonians where the 2ePCE correction model is given as a prefix \colorbox{red}{a} while the defining Hamiltonian matrix $\vec{h}^{\text{4c}}_{\rm def}$
is given as subscript $\colorbox{blue}{b}$, that is:
$\colorbox{red}{a}{\textsc{X2C}}_{\colorbox{blue}{b}}$.\\ 
In particular we have
\begin{equation*}
   \colorbox{red}{a} = 
   \begin{cases}
        {\rm 1e}   & \textbf{if}\ \text{no\ 2ePCE\ corrections added: $
        \Delta\tilde{\vec{F}}^{\text{2c,2e}}_{}= \vec{0}$} \\
        {\rm amf}  & \textbf{if}\ \text{atomic--mean\ field\ 2ePCE\ corrections added: $
        \Delta\tilde{\vec{F}}^{\text{2c,2e}}_{}\simeq\Delta\tilde{\vec{F}}^{\text{2c,2e}}_{\bigoplus}$ (see\ line \ref{alg:amfx2c:xcenergy} in Alg.~\ref{alg:amfX2C})}\\
        {\rm eamf} & \textbf{if}\ \text{extended\ atomic--mean\ field\ 2ePCE\ corrections added: $
        \Delta\tilde{\vec{F}}^{\text{2c,2e}}_{}\simeq \Delta\tilde{\vec{F}}^{\text{2c,2e}}_{\bigoplus}$ (see\ line \ref{alg:eamfX2C:deltaF2c2e} in Alg.~\ref{alg:eamfX2C})}\\
        {\rm \textsf{AMFI}} & \textbf{if}\ \text{atomic--mean\ field\ first--order (DKH1) spin--orbit 2ePCE\ corrections added: see\ Refs.~\citenum{hess96}\ and \ \citenum{schimmel96}}\\
        {\rm mmf} & \textbf{if}\ \text{post-SCF molecular--mean\ field\ 2ePCE\ corrections added: see\ Ref.~\citenum{sikkema09}}
   \end{cases}
\end{equation*}
and 
\begin{equation*}
   \colorbox{blue}{b} = 
   \begin{cases}
        {\rm D}   & \textbf{if}\ \text{$\vec{U}$\ is\ evaluated\ from $\vec{h}^{\text{4c}}_{\rm def} \equiv  \vec{h}^{\text{4c}}$ where $\vec{h}^{\text{4c}}$ is the one-electron Dirac Hamiltonian} \\
        {\rm DC}  & \textbf{if}\ \text{$\vec{U}$\ is\ evaluated\ from\ $\vec{h}^{\text{4c}}_{\rm def} \equiv  \vec{h}^{\text{4c}} + \vec{F}^{\text{4c,2e}}_{\bigoplus}$ with\ \underline{C}oulomb\ integrals\ contributing\ to\ $\vec{F}^{\text{4c,2e}}_{\bigoplus}$}\\
        {\rm DCG} & \textbf{if}\ \text{$\vec{U}$\ is\ evaluated\ from\ $\vec{h}^{\text{4c}}_{\rm def} \equiv  \vec{h}^{\text{4c}} + \vec{F}^{\text{4c,2e}}_{\bigoplus}$ with\ \underline{C}oulomb--\underline{G}aunt\   integrals\ contributing\ to\ $\vec{F}^{\text{4c,2e}}_{\bigoplus}$}\\
        {\rm DCB} & \textbf{if}\ \text{$\vec{U}$\ is\ evaluated\ from\ $\vec{h}^{\text{4c}}_{\rm def} \equiv  \vec{h}^{\text{4c}} + \vec{F}^{\text{4c,2e}}_{\bigoplus}$ with\ \underline{C}oulomb--\underline{B}reit\  integrals\ contributing\ to\ $\vec{F}^{\text{4c,2e}}_{\bigoplus}$}.\\
   \end{cases}
\end{equation*}

\begin{figure}
\begin{algorithm}[H]
\caption{Pseudo-code highlighting the essential steps for the eamfX2C approach.}
  \label{alg:eamfX2C}
  \begin{algorithmic}[1]
  \State \mycm{Initialize the \textcolor{blue}{molecular} effective density matrices}
  \State $
        \vec{D}^{\text{4c}}_{\bigoplus}=\vec{0};
        \qquad\ 
        \tilde{\vec{D}}^{\text{2c}}_{\bigoplus}=\vec{0}
        $
  \ForAll{\texttt{unique atom types $K$\ $\in$ molecule}}
    \State \textsf{Let} $\{\mu, \nu\} \in {\rm atomic\ basis}\ K$
    \State \mycm{Solve the 4c SCF equation}
    \State          $
               \vec{F}^{\text{4c}}_{K}\vec{c}^{\text{4c}}_{K}
                =  
               \vec{c}^{\text{4c}}_{K}\epsilon^{\text{4c}}_{K}$
    \,\, \textsf{with}
    \,\,
            $ F^{\text{4c}}_{K,\mu\nu}
              = 
              \begin{cases}
                  F^{\text{4c,HF}}_{K,\mu\nu}[\vec{D}^{\text{4c}}_{K}] 
                     = 
                     h^{\text{4c}}_{K,\mu\nu}
                     +
                     \sum\limits_{\gamma, \delta \in K}
                     G_{\mu\nu,\gamma\delta}^{\text{4c}}D^{\text{4c}}_{K,\delta\gamma}
                     \\[0.2cm]
                     F^{\text{4c,KS}}_{K,\mu\nu}[\vec{D}^{\text{4c}}_{K}] 
                     = 
                     h^{\text{4c}}_{K,\mu\nu}
                     +
                     \sum\limits_{\gamma, \delta \in K}
                     G^{\omega,\text{4c}}_{\mu\nu,\gamma\delta}D^{\text{4c}}_{K,\delta\gamma}
                     +
                     F^{\text{4c,xc}}_{K,\mu\nu}[\vec{D}^{\text{4c}}_{K}]
              \end{cases}
             $

        \State \mycm{Evaluate the \textcolor{blue}{atomic} X2C decoupling matrix $\vec{U}_{K}$ from  $\vec{F}^{\text{4c}}_{K}$ and calculate}
        \State
        $
        \tilde{\vec{D}}^{\text{2c}}_{K}
               =
               \left[ \vec{U}^{\dagger}_{K} \vec{D}^{\text{4c}}_{K} \vec{U}_{K} \right]^{\text{LL}}
        $
        
        \State \mycm{Add K-th \textcolor{blue}{atomic} effective density matrices $\vec{D}^{\text{4c}}_{K}$\ and $\tilde{\vec{D}}^{\text{2c}}_{K}$\ to the \textcolor{blue}{molecular} block}
        
        \State \label{alg:eamfX2c:4cDensblocked} $
               \vec{D}^{\text{4c}}_{\bigoplus} \gets  \vec{D}^{\text{4c}}_{K}; \qquad\ 
               \tilde{\vec{D}}^{\text{2c}}_{\bigoplus} \gets  \tilde{\vec{D}}^{\text{2c}}_{K}
               $
    \EndFor
    \State \textsf{Let} $\{\mu, \nu\} \in \text{full molecular basis}$
    \State \mycm{Evaluate the \textcolor{blue}{molecular} 4c 2e Fock matrix $\vec{F}^{\text{4c,2e}}_{\bigoplus}$ with elements}
   \State \label{alg:eamfX2C:line4cF}
   $
      {F}^{\text{4c,2e}}_{\bigoplus,\mu\nu} 
      =
      \begin{cases}
        F^{\text{4c,2e,HF}}_{\bigoplus,\mu\nu}[\vec{D}^{\text{4c}}_{\bigoplus}] 
        = 
        \sum\limits_{\gamma,\delta}
        G_{\mu\nu,\gamma\delta}^{\text{4c}}D^{\text{4c}}_{\bigoplus,\delta\gamma}
        
        \\[0.2cm]
        F^{\text{4c,2e,KS}}_{\bigoplus,\mu\nu}[\vec{D}^{\text{4c}}_{\bigoplus}] 
        = 
        \sum\limits_{\gamma, \delta}
        G^{\omega,\text{4c}}_{\mu\nu,\gamma\delta}D^{\text{4c}}_{\bigoplus,\delta\gamma}
        +
        F^{\text{4c,xc}}_{\mu\nu}[\vec{D}^{\text{4c}}_{\bigoplus}]
        \end{cases}
   $
   \State \mycm{If DFT, evaluate also the \textcolor{blue}{molecular} xc energy 
                ${E}_{xc}^{\text{4c}}[\vec{D}_{\bigoplus}^{\text{4c}}]$}
        
    \State \mycm{Evaluate the \textcolor{blue}{molecular} X2C decoupling matrix $\vec{U}$ from}
    \State
    $
      \tilde{\vec{h}}^{\text{4c}} 
      = 
      \vec{h}^{\text{4c}} 
      + 
      \vec{F}^{\text{4c,2e}}_{\bigoplus}
    $

     \State \mycm{Determine the \textcolor{blue}{molecular} 2e picture-change transformation correction as}
     \State\label{alg:eamfX2C:deltaF2c2e}
     $
      \Delta\tilde{\vec{F}}^{\text{2c,2e}}_{\bigoplus}
      =
      \left[ \vec{U}^{\dagger} \vec{F}^{\text{4c,2e}}_{\bigoplus} \vec{U} \right]^{\text{LL}}
      -
      \vec{F}^{\text{2c,2e}}
    $
    \State \textsf{where}
    \State
    $
      F^{\text{2c,2e}}_{\mu\nu}
      =
      \begin{cases}
         F^{\text{2c,2e,HF}}_{\mu\nu}[\tilde{\vec{D}}^{\text{2c}}_{\bigoplus}] 
         = 
         \sum\limits_{\gamma, \delta}
         G_{\mu\nu,\gamma\delta}^{\text{2c}}\tilde{D}^{\text{2c}}_{\bigoplus,\delta\gamma}
         \\[0.2cm]
         F^{\text{2c,2e,KS}}_{\mu\nu}[\tilde{\vec{D}}^{\text{2c}}_{\bigoplus}] 
         = 
         \sum\limits_{\gamma, \delta}
         G^{\omega,\text{2c}}_{\mu\nu,\gamma\delta}\tilde{D}^{\text{2c}}_{\bigoplus,\delta\gamma}
         +
         F^{\text{2c,xc}}_{\mu\nu}[\tilde{\vec{D}}^{\text{2c}}_{\bigoplus}]
         \end{cases}
        $
        
    \State \mycm{If DFT, determine also the \textcolor{blue}{molecular} PCE correction to the xc energy as}
    \State \label{alg:eamfX2c:xcenergy}
        $
         \Delta\tilde{E}^{\text{2c}}_{xc,\bigoplus} 
         = 
         {E}_{xc}^{\text{4c}}[\vec{D}_{\bigoplus}^{\text{4c}}] 
         - 
         {E}_{xc}^{\text{2c}}[\tilde{\vec{D}}_{\bigoplus}^{\text{2c}}]
        $
    
    \State \mycm{Solve the 2c SCF equation with the eamfX2C Fock matrix operator}
   \State
   $
   \vec{F}^{\text{2c}}\vec{c}^{\text{2c}}
      =
      \vec{c}^{\text{2c}}\epsilon^{\text{2c}}
       \,\, \textsf{with} \,\,
      F^{\text{2c}}_{\mu\nu}
      \equiv
      F^{\text{eamfX2C}}_{\mu\nu}
      =
      \underbrace{
      \left[ \vec{U}^{\dagger} \vec{h}^{\text{4c}} \vec{U} \right]^{\text{LL}}_{\mu\nu}
      +
      \Delta\tilde{F}^{\text{2c,2e}}_{\bigoplus,\mu\nu}
      }_{\text{static terms}}
      +
      \underbrace{
      F^{\text{2c,2e}}_{\mu\nu}[\vec{D}^{\text{2c}}]
      }_{\text{dynamic term}}
    $
   \end{algorithmic}
\end{algorithm}
\end{figure}

\section{Computational Details}\label{sec:compdet}

If not stated otherwise, all calculations reported in this work have been carried out by both \textsc{Dirac}~\cite{dirac-paper} and \textsc{ReSpect}~\cite{ReSpect2020} programs, making use of a common computational setup: (i) a finite value for the speed of 
light $c=137.03599907400\ a_0E_h/\hbar$,~\cite{Mohr2008} (ii) a point nucleus model for all atomic nuclei to ease comparison between 
data obtained by the programs, (iii) an explicit inclusion of $(SS|SS)$-type electron repulsion AO-integrals, (iv) atom-centered uncontracted 
Gaussian-type basis sets of double-$\zeta$ quality (dyall.v2z, dubbed \texttt{v2z}) for each unique atom type,~\cite{Dyall_Basis_2p-3p,Dyall_Basis_4p-6p,Dyall_Basis_4p-6p_rev,Dyall_Basis_7p,Dyall_Basis_5d,Dyall_Basis_5d_rev,Dyall_Basis_6d,Dyall_Basis_Source} 
(v) \textsc{Dirac}'s default numerical integration 
grids consisting of the basis-set adaptive radial quadrature by Lindh~\emph{et al.},~\cite{Lindh2001} and the angular quadrature by 
Lebedev~\cite{Lebedev1975,Lebedev1976,Lebedev1977} (to achieve consistent exchange--correlation PCE corrections 
by both programs, it turned out be crucial to use integration grids of identical composition and quality)\ and (vi) a threshold for SCF convergence of $10^{-7}$ in the DIIS\cite{Pulay_CPL1980}  error vector. 
All atomic and molecular calculations with \textsc{Dirac} were performed within a Kramers-restricted (KR) formalism, 
employing for open-shell systems either an average-of-configuration (AOC) approach~\cite{Thyssen_phd2004} (HF) or a fractional 
occupation (FO) approach (KS-DFT). In the case of group-16 diatomics (chalcogenide series), AOC HF calculations take into account 
all possible configurations of six electrons in 8 Kramers-paired spinors (i.e., representing the $\pi,\pi^\ast$\ valence shells). 
In \textsc{ReSpect}, molecular open-shell calculations were performed within a Kramers-unrestricted (KU) formalism,~\cite{ReSpect2020} whereas atomic results were obtained with the KR FO approach, both for HF and KS-DFT calculations.
All KS-DFT calculations were carried out with either a PBE or PBE0 exchange--correlation functional.~\cite{Perdew1996-1,Perdew1996-2,Adamo1998} 

For the lighter noble gas dimers, internuclear distances were taken from experimentally available data \cite{ogil92}\ whereas for the heavier homologues Rn$_2$\ and Og$_2$, respectively, computationally optimized structures were taken from Ref.~\citenum{shee15}. Similarly, in the case of the chalcogene series, all geometries were taken from Ref.~\citenum{rota:114106}, except for the heaviest diatomic system Lv$_2$ for which the internuclear distance of $R_e=3.230$\ \AA\ was extracted by visual inspection from Figure 1 of Ref.~\citenum{trom19}. 
Table \ref{tab:group-X-struct}\ summarizes the structural parameters for all group 16 and group 18 diatomics employed in this work. 

\begin{table}
\caption{Structural parameters of the group 16 (left-hand side) and group 18 diatomics (right-hand side) considered 
in this work. All internuclear distances are given in \AA.\label{tab:group-X-struct}}
 \begin{tabular}{l*{2}r@{\hskip 0.5cm}l*{2}r}
 \hline \hline\noalign{\smallskip}
 molecule & $r_{\rm X-X}$\ & Reference & molecule & $r_{\rm X-X}$\ & Reference \\ \hline

&& & He$_2$ & 2.970   & \citenum{ogil92} \\
O$_2$  & 1.20752 & \citenum{rota:114106} & Ne$_2$ & 3.091   & \citenum{ogil92} \\
S$_2$  & 1.889   & \citenum{rota:114106} & Ar$_2$ & 3.756   & \citenum{ogil92} \\
Se$_2$ & 2.166   & \citenum{rota:114106} & Kr$_2$ & 4.008   & \citenum{ogil92} \\
Te$_2$ & 2.557   & \citenum{rota:114106} & Xe$_2$ & 4.363   & \citenum{ogil92} \\
Po$_2$ & 2.795   & \citenum{rota:114106} & Rn$_2$ & 4.427   & \citenum{shee15}\\
Lv$_2$ & 3.230   & \citenum{trom19}      & Og$_2$ & 4.329   & \citenum{shee15} \\ \hline 
\end{tabular}
\end{table}

In the case of the methane molecule CH$_4$\ discussed in Section \ref{sec:CH4-ultra}, we assumed a T$_{d}$-symmetrical molecular framework with a C-H internuclear distance of 1.091 \AA\ and a $<$H-C-H bond angle of 109.471 degrees. In order to enhance relativistic effects, we scaled down the speed of light $c$\ by a factor of 10, corresponding to an actual value of $c_{\rm scaled}=13.703599907400\ a_0E_h/\hbar$, for both the atomic as well as the molecular calculations.

The absolute contact densities and contact density shifts for selected (closed-shell) copernicium fluorides (CnF$_n$, $n=0,2,4,6$), discussed in Section \ref{sec:group12-fluorides}, were calculated from mean-field HF wave functions employing a 4c Dirac-Coulomb as well as the X2C Hamiltonian supplemented with various 2ePCE corrections. The structures for each of the copernicium fluorides were optimized within a 4c Dirac-Coulomb framework by means of KS-DFT calculations employing the PBE0 exchange-correlation
functional. Following the very recent work of Hu and Zou,\cite{hush21} we assumed for the structure optimization a linear ($n=2$), square-planar ($n=4$) and octahedral ($n=6$) geometry for the respective copernicium fluorides CnF$_n$. The resulting equilibrium Cn-F internuclear distances are compiled in Table \ref{tab:cnfn-structures} along with the corresponding double-group symmetry. 
It is worthwhile to note that, given that the primary concern of our the present study is not to provide an accurate computation of the contact density with respect to a converged basis set saturation at the heavy nuclei Cn, we did not further pursue any further
augmentation of the set of primitives in the basis set as, for example, done in our earlier works in Refs. \citenum{knec11a} and \citenum{Knecht2012103}. 

Finally, Section \ref{sec:group17-anion}\ comprises an assessment of the accuracy and suitability of various 2c approaches to adequately describe (absolute) K- and L-edge core-ionization energies as well as  L$_3$-L$_2$\ edge spin-orbit splittings, denoted as $\Delta^{\rm SO}_{\rm L}$, for heavy $d$-\ and $p$-block compounds. To this end, we considered one atomic anion (At$^-$) and two anionic and neutral molecular cases, respectively. In the former case we employed the same computational setup for the SCF and EOM-CCSD\cite{Shee:JCP2018}  calculations as described in full detail in Ref.~\citenum{halb21}\ which provides high-quality computational reference data. 
In the remaining molecular examples, we employed for CnF$_6$\ the optimized molecular structure listed in Table \ref{tab:cnfn-structures}, while for [Au(Cl)$_4$]$^{-}$\ the optimized molecular structure has been taken from Table 1 (column MP2/aug-cc-pVTZ) in Ref.~\citenum{harg01a}. In either molecular cases, we correlated for the EOM-CCSD step all electrons and introduced an energy-based cutoff in the virtual spinor space at 3 Hartree making use of the dyall.v2z basis sets for all atom types. Since we are solely interested in a genuine comparison of different two-component Hamiltonian models rather than achieving \textit{quantitatively converged} results for the K-, L-edge and M-edge (the latter only for the [Au]-complex) ionisation potentials which would call for, for example, to make use of tailored basis sets \cite{ambr21a}, the latter motivates for the present work our choice to merely aim at a \textit{qualitative} electron correlation treatment.

\section{Results and discussion}\label{sec:results}

In this section we will critically assess the accuracy of our newly developed 2ePCE correction approaches for all-electron X2C HF and DFT calculations in the two major, common use cases, that is (i) with a variational account of SO interaction as well as (ii) in a genuine spin-free SC framework. A detailed summary of the notation of the 2ePCE correction applied to the X2C Hamiltonian can be found in Section \ref{sec:notations}.

In Section \ref{sec:core-orbital-energies}, we commence with a discussion of the spinor energies of Og$_2$, a prototypical, closed-shell superheavy diatomic molecule, optimized both within a mean-field HF and a KS-DFT computational model (Section \ref{sec:core-orbital-energies-HF}). Results for the lighter homologues of the corresponding group 18 diatomics can be found in the public research repository ZENODO (see Section~\ref{sec:SIData} for more details). Along the same lines and as an example of an open-shell diatomic molecule, we consider in Section \ref{sec:core-orbital-energies-Te2}\ the case of Te$_2$\ as a representative of the group-16 diatomics. Results for the remaining group-16 diatomics listed in Table \ref{tab:group-X-struct}\ can also be found in the ZENODO repository (see Section~\ref{sec:SIData} for more details). To conclude the discussion on total as well as spinor energies, we assess in Section \ref{sec:CH4-ultra}\ the numerical performance of our selection of PCE-corrected X2C models for the case of an ``ultrarelativistic" methane molecule employing a ten-fold reduced speed of light $c$, that is $c/10$. 

Next, in Section \ref{sec:group12-fluorides}\ we evaluate the suitability of our (e)amf-X2C models for the calculation of absolute contact densities at a heavy nuclear center and, equally important, for contact density \textit{shifts}. The latter play, for example, an important role in computational models for the determination of isomer shifts that are accessible in experimental M\"ossbauer spectroscopy. 
To this end, we perform contact density calculations for a series $n$ (with $n=0,2,4,6$) of fluoride compounds of the heaviest group 12 member Cn, ranging from the bare Cn atom to the hexafluoride CnF$_6$.

In Section \ref{sec:group17-anion}\ we conclude our assessment by focusing on the calculation of correlated X-ray core ionization energies. Besides the At$^{-}$\ mono-anion for which benchmark data is available in the literature,\cite{halb21} we consider two molecular applications of $5d$\ and $6d$-containing molecules by taking advantage of the recently developed EOM-CCSD approach for core spectroscopy.\cite{halb21}

\subsection{Spinor energies of (super)heavy diatomic molecules}\label{sec:core-orbital-energies}

\subsubsection{Closed-shell Og$_2$}\label{sec:core-orbital-energies-HF}

\begin{figure}
    \centering
\begin{subfigure}{0.5\textwidth}
  \centering
  \includegraphics[width=1.0\linewidth]{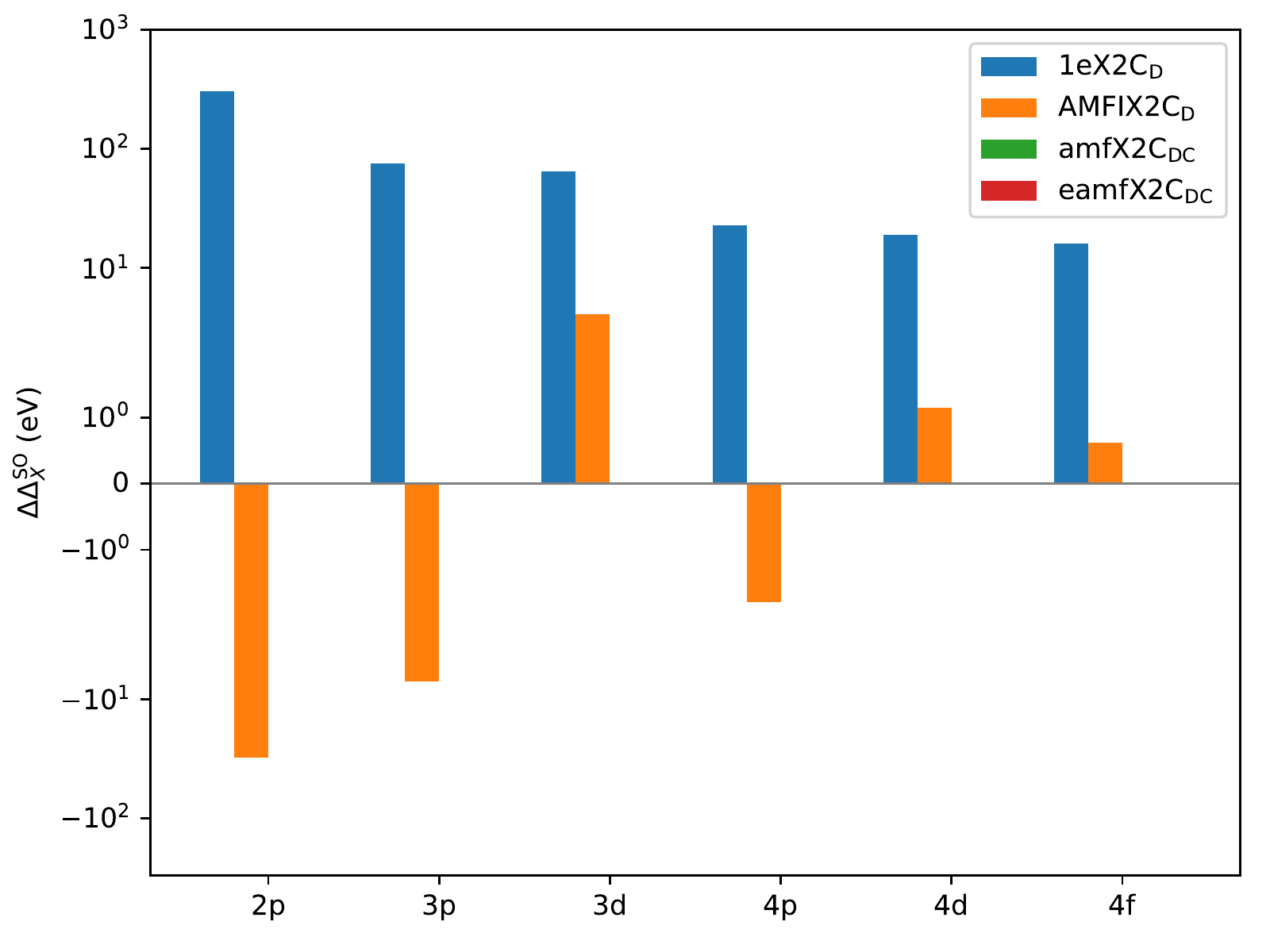}
  \caption{HF}
  \label{fig:og2-SOsplits-hf}
\end{subfigure}%
\begin{subfigure}{0.5\textwidth}
  \centering
  \includegraphics[width=1.0\linewidth]{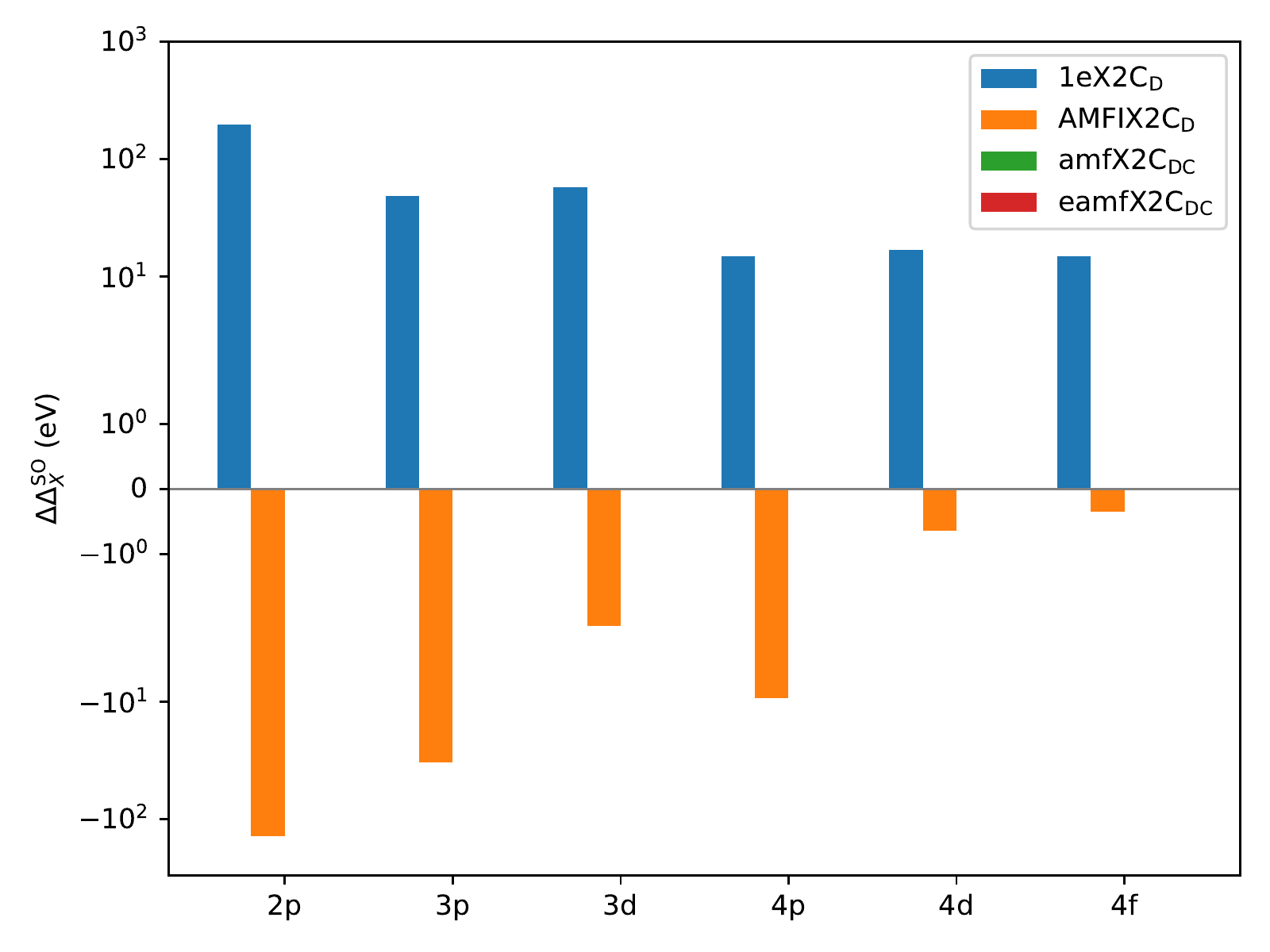}
  \caption{DFT/PBE}
  \label{fig:og2-SOsplits-pbe}
\end{subfigure}
\caption{Differences of spin-orbit splittings ($\Delta^{\text{SO}}_{X}$) of the inner-core to outer-core Og-atomic-like shells in Og$_2$\ with respect to the $^4$DC reference values either within a HF approach (panel (a)) or a DFT/PBE approach (panel (b)). All data is compiled from the SCF spinor energies listed in Tables \ref{tab:og2-HF} and \ref{tab:og2-DFT}, respectively. Energy differences are given in eV. Note that errors associated with the (e)amfX2C models are not visible in the figures.}
    \label{fig:og2-SOsplits}
\end{figure}

In the following, we will assess the numerical performance of our atomic mean-field PCE correction model and its extended version within the context of an exact two-component decoupling approach by considering as prime example the heaviest group-18 dimer, namely Og$_2$. Since the molecule is closed-shell in its electronic ground state, both the Kramers-restricted and the Kramers-unrestricted SCF  formalism implemented in \textsc{DIRAC} and \textsc{ReSpect}, respectively, converge to the same solution. In order to underline the importance of a simultaneous treatment of 2eSC and 2eSO PCE corrections within the X2C Hamiltonian framework, we compile in Table \ref{tab:og2-HF} a selected set of HF spinor energies for Og$_2$, ranging from the inner- to outer-core as well as to the valence region, and compare the various X2C-based spinor energies with the 4c Dirac-Coulomb reference data ($^4$DC; sixth column in Table \ref{tab:og2-HF}). In addition, the left panel of Figure \ref{fig:og2-SOsplits} comprises the HF-based deviations for SO splittings of the inner-core and outer-core shells of Og$_2$ with predominant atomic-like character illustrated for results obtained with the various two-component Hamiltonian schemes listed in Table \ref{tab:og2-HF} by  comparison to the $^4$DC reference. Finally, the right panel of Figure  \ref{fig:og2-SOsplits}\ provides a similar comparison for a correlated KS-DFT-based approach employing the PBE functional where the underlying  absolute energies are summarized in Table \ref{tab:og2-DFT}. 

\paragraph{HF}

\begin{table}
 \caption{SCF total energy ($E$) and spinor energies of selected doubly-degenerate occupied spinors ($\epsilon$) for Og$_{2}$ as obtained from HF/\texttt{v2z} calculations within a four-component Dirac-Coulomb ({$^4$DC}) as well as a two-component Hamiltonian framework, including the new (e)amfX2C$_{\rm DC}$ models. All energies are given in Hartree.}
 \centering
 \begin{tabular}{l*{5}r}
 \hline \hline\noalign{\smallskip}
{}&                      {1eX2C$_{\rm D}$} & {\textsf{AMFI}X2C$_{\rm D}$}&       {amfX2C$_{\rm DC}$}&      {eamfX2C$_{\rm DC}$}&           {$^4$DC}\\\hline
$E$&  -110045.25693&  -110015.96688&  -110116.09102&  -110116.09102&  -110116.09101\\
$\epsilon_{1-2}$&    -8248.36274&    -8248.69505&    -8272.12530&    -8272.12529&    -8272.12529\\
$\epsilon_{3-4}$&    -1733.89154&    -1734.00101&    -1738.99764&    -1738.99763&    -1738.99763\\
$\epsilon_{5-6}$&    -1693.29607&    -1683.36133&    -1686.06374&    -1686.06374&    -1686.06374\\
$\epsilon_{7-10}$&    -1133.93651&    -1136.41886&    -1137.97905&    -1137.97904&    -1137.97904\\
$\epsilon_{11-12}$&     -474.97349&     -475.01315&     -476.18010&     -476.18010&     -476.18010\\
$\epsilon_{13-14}$&     -454.67004&     -452.30145&     -452.93331&     -452.93331&     -452.93331\\
$\epsilon_{15-18}$&     -317.10573&     -317.76956&     -318.14142&     -318.14142&     -318.14142\\
$\epsilon_{19-22}$&     -287.84702&     -286.46016&     -286.46862&     -286.46861&     -286.46861\\
$\epsilon_{23-28}$&     -264.51100&     -265.35539&     -265.51476&     -265.51476&     -265.51476\\
$\epsilon_{29-30}$&     -142.09699&     -142.11136&     -142.43246&     -142.43246&     -142.43246\\
$\epsilon_{31-32}$&     -131.90002&     -131.20172&     -131.36462&     -131.36462&     -131.36462\\
$\epsilon_{33-36}$&      -91.64727&      -91.85115&      -91.94818&      -91.94818&      -91.94818\\
$\epsilon_{37-40}$&      -76.61348&      -76.20853&      -76.19682&      -76.19682&      -76.19682\\
$\epsilon_{41-46}$&      -70.00892&      -70.25753&      -70.28799&      -70.28799&      -70.28799\\
$\epsilon_{47-52}$&      -50.08877&      -49.76060&      -49.73704&      -49.73703&      -49.73703\\
$\epsilon_{53-60}$&      -47.74819&      -47.99085&      -47.99004&      -47.99004&      -47.99004\\
$\ldots$&               $\ldots$&        $\ldots$&       $\ldots$&       $\ldots$&       $\ldots$\\
$\epsilon_{110}$&       -1.47090&       -1.48271&       -1.48161&       -1.48162&       -1.48162\\
$\epsilon_{111}$&       -1.31383&       -1.31314&       -1.31699&       -1.31698&       -1.31698\\
$\epsilon_{112}$&       -1.31254&       -1.31185&       -1.31572&       -1.31571&       -1.31571\\
$\epsilon_{113}$&       -0.74647&       -0.73730&       -0.73819&       -0.73819&       -0.73819\\
$\epsilon_{114}$&       -0.74381&       -0.73455&       -0.73545&       -0.73545&       -0.73545\\
$\epsilon_{115}$&       -0.31691&       -0.31826&       -0.31821&       -0.31822&       -0.31822\\
$\epsilon_{116}$&       -0.30372&       -0.30516&       -0.30512&       -0.30512&       -0.30512\\
$\epsilon_{117}$&       -0.29260&       -0.29413&       -0.29411&       -0.29411&       -0.29411\\
$\epsilon_{118}$&       -0.28036&       -0.28196&       -0.28194&       -0.28193&       -0.28193\\
 \hline \hline
\end{tabular}
\label{tab:og2-HF}
\end{table}

In line with previous works,\cite{vanW05a,peng12a} we find the largest deviations within an X2C framework from the reference 4c spinor energies in an HF approach for the innermost $s$\ and $p$\ shells where 2eSO ($p$\ shells) and 2eSC PCE corrections ($s$\ and $p$\ shells) are expected to be of utmost importance (see also the discussion of core-ionization energies in Section \ref{sec:group17-anion}). Hence, considering first the bare one-electron X2C (second column, {{1eX2C$_{\rm D}$}}\ in Table \ref{tab:og2-HF}), which ignores 2e picture changes altogether, we encounter deviations up to +23.8 Hartree with respect to the four-component reference data for the innermost $s$\ shells and up to -7.2 Hartree for the lowest-lying $p$\ shells. Next, by taking into account atomic SO mean-field PCE corrections within the AMFI model (third column, {\textsf{AMFI}X2C$_{\rm D}$}) results in a minor improvement of about -0.4 Hartree for the inner $s$\ shells while the lowest-lying $p$\ shells\ become destabilized through the PCE corrections by about +10 Hartree leading to a deviation of $\approx\ +2.7$\ Hartree wrt the corresponding 4c reference values.

By contrast, both our amfX2C$_{\rm DC}$ and eamfX2C$_{\rm DC}$ PCE correction schemes for the X2C Hamiltonian yield spinor energies which merely differ by 10 $\mu$Hartree or less for the innermost $s$\ shells -- and likewise for the $p$\ shells -- of Og$_2$ from the 4c reference data. These findings strikingly illustrate the excellent numerical performance of our newly proposed amf-based 2eSC- and 2eSO-PCE corrections applied in a molecular framework.  
Moreover, in particular in the core region close to a (heavy) nucleus, SO splittings are a crucial measure since they probe the ability of PCE-corrected 2c schemes to provide \textit{quantitative relative} energies. Here, calculations employing the 1eX2C$_{rm D}$ as well as the \textsf{AMFI}X2C$_{\rm D}$ Hamiltonian yield SO-splittings for the atomic-like shells ($\Delta^{\mbox{SO}}_{X}, X=p,d,f$\; obtained as energy difference $\epsilon_{X_{(2l+1)/2}}-\epsilon_{X_{(2l-1)/2}}$) in Og$_{2}$ which deviate significantly from the $^4$DC reference data as illustrated in Figure \ref{fig:og2-SOsplits-hf}\ with data obtained from Table \ref{tab:og2-HF}. For example, for the bare 1eX2C$_{\rm D}$ approach we find deviations in $\Delta^{\mbox{SO}}_{X}$\ of up to $\approx$ +11.3 Hartree\ for the $2p$\ shell\ which corresponds to an overestimation of the splitting by $\approx$\ 2\%. Moving to outer-core shells, the overestimation of the SO splitting $\Delta^{\mbox{SO}}$\ becomes even worse with deviations as large as  $\approx$\ +25\% for $\Delta^{\mbox{SO}}_{4f}$. As can be seen from Figure  \ref{fig:og2-SOsplits-hf}, the latter deviations can be reduced significantly for all inner- and outer-core spin-orbit-split shells through the introduction of AMFI-based SO mean-field PCE corrections within \textsf{AMFI}X2C$_{\rm D}$. Finally, as it is evident from the matching \textit{absoulute}\ spinor energies discussed above, all SO splittings considered in Figure  \ref{fig:og2-SOsplits-hf} obtained within our (e)amfX2C$_{\rm DC}$ Hamiltonian frameworks match (within significant digits) their 4c reference data (errors are therefore not visible in the Figure), underlining once more the importance of taking into account both 2eSC- and 2eSO-PCE corrections in an X2C many-electron Hamiltonian framework. 

In passing we note that the numerical performance of our (e)amfX2C models not only holds for the inner- and outer-core but also for the correspondings valence shells ($\epsilon_{110-118}$\ in Table \ref{tab:og2-HF}) of the diatomic Og$_2$\ where the $^4$DC and (e)amfX2C$_{\rm DC}$ data essentially remains indistinguishable within significant digits. Noteworthy, in the (outer-)valence region, the \textsf{AMFI}X2C$_{\rm D}$ approach leads to  absolute spinor energies which differ by less than 10$^{-3}$\ Hartree from their reference values. Hence, the latter may explain why this PCE correction scheme has successfully been applied in the past in numerous numerical applications that particularly probed valence-dominated properties. Finally, our data in Table \ref{tab:og2-HF} further shows that neglecting PCE corrections at all results even for valence spinors in absolute errors for the spinor energies on the order of 10$^{-2}$\ Hartree. 

\paragraph{DFT/PBE}

What about the numerical performance of our (e)amf PCE correction models in a correlated framework? To this end, we consider in the following the same prime superheavy diatomic molecular system Og$_2$\ (\textit{vide supra}) within a DFT/PBE-based SCF approach. A particularity of our (e)amf PCE correction models is rooted in the fact that, as illustrated in Algorithms \ref{alg:amfX2C}\ and \ref{alg:eamfX2C}, respectively, both models enable not only a basis-set dependent but also a \textit{self-consistent-field model dependent}\ PCE correction\ which originate from the specific contributions that enter the corresponding 2e Fock matrices. The latter implies that our (e)amf PCE correction models provide tailor-made PCE corrections which explicitly account for the subtleties that arise from the employed exchange-correlation functional within a KS-DFT-based SCF approach. By contrast, to the best of our knowledge common PCE schemes such as the AMFI approach do -- by construction -- \textit{not}\ allow to distinguish between 2eSO PCE corrections for the X2C Hamiltonian that either aim for an ensuing (uncorrelated) 2c HF or (correlated) KS-DFT-based many-electron SCF calculation. Bearing these subtle, yet crucial details in mind, the strikingly excellent numerical performance of our (e)amfX2C$_{\rm DC}$ models wrt the $^4$DC reference spinor energies as well as total energies which are illustrated in Table \ref{tab:og2-DFT}\ not only underlines the outstanding numerical performance of our newly proposed PCE correction \textit{ans\"atze} but is also in perfect agreement with our previous conclusions within the HF approach (\textit{vide supra}). Moreover, the SO splittings $\Delta^{\text{SO}}_{X}$\ of the (e)amfX2C$_{\rm DC}$ and $^4$DC\ cases match again exactly within significant digits for all the selected inner-core and outer-core atomic-like shells shown in Figure \ref{fig:og2-SOsplits-pbe}. Notably, as indicated above, the (basis-set dependent) AMFI-based SO PCE corrections are SCF-model independent and, hence, strictly identical for both common use cases, \textit{viz.} in an X2C-HF and X2C-KS-DFT approach. Consequently, AMFI does not include \textit{a priori}\ any PCE corrections on the SO splitting originating from amf \textit{two-electron correlation effects} which should primarily have an impact on the resulting splitting of the most strongly SO-split $p$\ shells. A close inspection of the left (HF) and right (DFT/PBE) panels of Figure \ref{fig:og2-SOsplits}\ reveals that the deviations from the $^4$DC reference for $\Delta^{\text{SO}}_{X} (X=2p,3p,4p)$\ are indeed  systematically larger in the (correlated) DFT/PBE case. 

\begin{table}
\caption{SCF total energy ($E$) and spinor energies ($\epsilon$) of the doubly-degenerate occupied spinors for Og$_2$ as obtained from  DFT/PBE/\texttt{v2z} calculations within a four-component Dirac-Coulomb ({$^4$DC}) as well as a two-component Hamiltonian framework, including the new (e)amfX2C$_{\rm DC}$ models. All energies are given in Hartree.}
 \begin{tabular}{l*{5}r}
\hline \hline \noalign{\smallskip}
{} & {1eX2C$_{\rm D}$} & {\textsf{AMFI}X2C$_{\rm D}$} & {amfX2C$_{\rm DC}$} & {eamfX2C$_{\rm DC}$} & {$^4$DC} \\ \hline
$E$&  -110101.19289&  -110071.81703&  -110191.68717&  -110191.68717&  -110191.68716\\
$\epsilon_{1-2}$&    -8194.40021&    -8194.74358&    -8228.57826&    -8228.57826&    -8228.57826\\
$\epsilon_{3-4}$&    -1714.66379&    -1714.76764&    -1720.61964&    -1720.61964&    -1720.61964\\
$\epsilon_{5-6}$&    -1675.00848&    -1665.13958&    -1672.43570&    -1672.43570&    -1672.43570\\
$\epsilon_{7-10}$&    -1119.86023&    -1122.32402&    -1124.50368&    -1124.50367&    -1124.50367\\
$\epsilon_{11-12}$&     -465.42115&     -465.45560&     -466.74289&     -466.74289&     -466.74289\\
$\epsilon_{13-14}$&     -445.52684&     -443.18339&     -444.87850&     -444.87850&     -444.87849\\
$\epsilon_{15-18}$&     -309.80499&     -310.45933&     -310.94433&     -310.94433&     -310.94433\\
$\epsilon_{19-22}$&     -281.43839&     -280.06878&     -280.35730&     -280.35730&     -280.35730\\
$\epsilon_{23-28}$&     -258.40460&     -259.23760&     -259.44279&     -259.44279&     -259.44279\\
$\epsilon_{29-30}$&     -137.01495&     -137.02661&     -137.36644&     -137.36644&     -137.36644\\
$\epsilon_{31-21}$&     -127.09144&     -126.40362&     -126.86089&     -126.86089&     -126.86089\\
$\epsilon_{33-26}$&      -87.69123&      -87.88959&      -88.00419&      -88.00418&      -88.00418\\
$\epsilon_{37-40}$&      -73.27801&      -72.88107&      -72.93437&      -72.93437&      -72.93437\\
$\epsilon_{41-46}$&      -66.86798&      -67.11111&      -67.14079&      -67.14078&      -67.14078\\
$\epsilon_{47-52}$&      -47.78011&      -47.45864&      -47.45576&      -47.45576&      -47.45576\\
$\epsilon_{53-60}$&      -45.50042&      -45.73773&      -45.72233&      -45.72233&      -45.72233\\
$\ldots$ & $\ldots$ & $\ldots$ & $\ldots$ & $\ldots$ & $\ldots$ \\
$\epsilon_{110}$&       -1.16575&       -1.17660&       -1.17408&       -1.17409&       -1.17409\\
$\epsilon_{111}$&       -1.00557&       -1.00535&       -1.00795&       -1.00795&       -1.00795\\
$\epsilon_{112}$&       -1.00485&       -1.00463&       -1.00724&       -1.00724&       -1.00724\\
$\epsilon_{113}$&       -0.54122&       -0.53307&       -0.53604&       -0.53603&       -0.53603\\
$\epsilon_{114}$&       -0.53907&       -0.53085&       -0.53384&       -0.53384&       -0.53384\\
$\epsilon_{115}$&       -0.20929&       -0.21028&       -0.21007&       -0.21007&       -0.21007\\
$\epsilon_{116}$&       -0.19942&       -0.20048&       -0.20027&       -0.20027&       -0.20027\\
$\epsilon_{117}$&       -0.19101&       -0.19215&       -0.19194&       -0.19194&       -0.19194\\
$\epsilon_{118}$&       -0.18304&       -0.18421&       -0.18401&       -0.18400&       -0.18400\\
\hline \hline
\end{tabular}
\label{tab:og2-DFT}
\end{table}

\paragraph{On the importance of two-electron scalar-relativistic PCE corrections}

In the previous paragraphs, we discussed the performance of our newly proposed (e)amf PCE corrections for the X2C Hamiltonian in either a HF or KS-DFT framework with a particular focus on \textit{relative} spinor energies of the superheavy diatomic molecule Og$_2$, that is, for example on the resulting SO splittings of inner- and outer-core atomic-like shells by comparison to the corresponding $^4$DC reference data. In order to highlight the full potential of our (e)amf PCE models, let us recall that our 2ePCE correction models take into account both 2eSO \textit{and} 2eSC correction terms. Whereas 2eSO PCE corrections are common to include in an (exact) two-component Hamiltonian framework for many-electron systems,\cite{liuj18,hess96,schimmel96,vanW05a} the inclusion of 2eSC-PCE correction terms is less so, despite their apparent significance to be illustrated in the following. To this end, we turn to a genuine spinfree SC framework by eliminating all spin-dependent terms from the parent $^4$DC Hamiltonian by means of the Dirac relation.\cite{dyall94,visscher:jcp2000} Hence, results obtained on the basis of the SC-$^4$DC Hamiltonian will serve as references for the discussion of the numerical performance of various PCE-corrected SC-X2C Hamiltonian models. For the ease of comparison with the above spin-dependent data, we consider in Tables \ref{tab:og2-sc-HF} and \ref{tab:og2-sc-DFT}, respectively, in a spinfree \textit{ansatz} the same superheavy diatomic molecule Og$_2$. 

\begin{table}
 \caption{SCF total energy ($E$) and orbital energies of selected doubly-degenerate occupied orbitals ($\epsilon$) for Og$_{2}$ as obtained from HF/\texttt{v2z} calculations within a scalar-relativistic (SC) four-component Dirac-Coulomb ({$^4$DC}) as well as a two-component Hamiltonian framework, including the new SC-(e)amfX2C$_{\rm DC}$ models. All energies are given in Hartree.}
 \label{tab:og2-sc-HF}
 \centering
 \begin{tabular}{l*{5}r}
 \hline \hline\noalign{\smallskip}
{} & {SC-1eX2C$_{\rm D}$} & {SC-\textsf{AMFI}X2C$_{\rm D}$} & {SC-amfX2C$_{\rm DC}$} & {SC-eamfX2C$_{\rm DC}$} & {SC-{$^4$DC}} \\ \hline
$E$&  -109086.48892&  -109086.48892&  -109171.75916&  -109171.75916&  -109171.75917\\
$\epsilon_{1-2}$&    -8263.96172&    -8263.96172&    -8291.15582&    -8291.15582&    -8291.15582\\
$\epsilon_{3-4}$&    -1738.62822&    -1738.62822&    -1743.94621&    -1743.94621&    -1743.94621\\
$\epsilon_{5-10}$&    -1263.67816&    -1263.67816&    -1264.70996&    -1264.70996&    -1264.70996\\
$\epsilon_{11-12}$&     -476.53515&     -476.53515&     -477.77497&     -477.77497&     -477.77497\\
$\epsilon_{13-18}$&     -350.21316&     -350.21316&     -350.49958&     -350.49958&     -350.49958\\
$\epsilon_{19-28}$&     -274.23599&     -274.23599&     -274.32251&     -274.32250&     -274.32250\\
$\epsilon_{29-30}$&     -142.64244&     -142.64244&     -142.98885&     -142.98885&     -142.98885\\
$\epsilon_{31-36}$&     -101.55738&     -101.55738&     -101.63035&     -101.63035&     -101.63035\\
$\epsilon_{37-46}$&      -72.82370&      -72.82370&      -72.83682&      -72.83682&      -72.83682\\
$\epsilon_{47-60}$&      -48.96807&      -48.96807&      -48.96163&      -48.96163&      -48.96163\\
$\ldots$ & $\ldots$ & $\ldots$ & $\ldots$ & $\ldots$ & $\ldots$ \\
$\epsilon_{110}$&       -1.61633&       -1.61633&       -1.61483&       -1.61482&       -1.61482\\
$\epsilon_{111}$&       -1.31132&       -1.31132&       -1.31593&       -1.31593&       -1.31593\\
$\epsilon_{112}$&       -1.31005&       -1.31005&       -1.31467&       -1.31467&       -1.31467\\
$\epsilon_{113}$&       -0.41445&       -0.41445&       -0.41435&       -0.41435&       -0.41435\\
$\epsilon_{114}$&       -0.39648&       -0.39648&       -0.39639&       -0.39639&       -0.39639\\
$\epsilon_{115}$&       -0.39648&       -0.39648&       -0.39639&       -0.39639&       -0.39639\\
$\epsilon_{116}$&       -0.38981&       -0.38981&       -0.38972&       -0.38972&       -0.38972\\
$\epsilon_{117}$&       -0.38981&       -0.38981&       -0.38972&       -0.38972&       -0.38972\\
$\epsilon_{118}$&       -0.37349&       -0.37349&       -0.37341&       -0.37341&       -0.37341\\
\hline \hline
\end{tabular}
\end{table}

A close inspection of both tables first shows that the bare (no PCE corrections) SC-1eX2C$_{\rm D}$ and the SC-\textsf{AMFI}X2C$_{\rm D}$ Hamiltonians yield within either computational model, \textit{viz.} HF and DFT/PBE, strictly matching numerical results. The reason is that with the elimination of any spin-dependent term from the (parent) 4c Hamiltonian, the AMFI PCE corrections simply become zero. Moreover, as could be expected, the largest 2eSC PCE corrections are encountered for the inner $s$\ shells (molecular spinors $\epsilon_{1-4}$\ in Tables \ref{tab:og2-sc-HF} and \ref{tab:og2-sc-DFT})\ with deviations for SC-1eX2C$_{\rm D}$ ($\equiv$\ SC-AMFIX2C$_{\rm D}$) up to 27.2 Hartree in the HF and 35.3 Hartree in the DFT/PBE case compared to the SC-$^4$DC reference data. By moving to the outer-core and up to occupied molecular spinors close to the Fermi level, 2eSC PCEs start to fade significantly with absolute deviations for the HOMO and HOMO-1 amounting to less than $10^{-4}$\ Hartree. By contrast, our SC-(e)amfX2C$_{\rm DC}$ models provide an even higher numerical accuracy by at least one order of magnitude ($< 10^{-5}$) for \textit{all}\ occupied molecular spinors summarized in Tables \ref{tab:og2-sc-HF} and \ref{tab:og2-sc-DFT}, that is ranging from the innermost $s$\ shells to the Fermi level. The latter findings therefore unequivocally illustrate that our atomic SC-(e)amfX2C$_{\rm DC}$ PCE correction models are capable of efficiently correcting for 2ePCEs in a molecular framework. Consequently, this distinct asset of our (e)amfX2C models is a key ingredient for their above discussed numerical success in a spin-dependent Hamiltonian framework where 2eSC and 2eSO coupling contributions are both simultaneously at play\ and should not be considered on a different footing. In passing we further note that also in the present spinfree case the total SCF energies obtained within either our (e)amfX2C or a 4c Hamiltonian framework agree up to $\mu$-Hartree accuracy, regardless of the underlying SCF \textit{ansatz}.

\begin{table}
 \caption{SCF total energy ($E$) and spinor energies of selected doubly-degenerate occupied spinors ($\epsilon$) for Og$_{2}$ as obtained from  
 DFT/PBE/\texttt{v2z} calculations within a scalar-relativistic (SC) four-component Dirac-Coulomb ({$^4$DC}) as well as a two-component Hamiltonian framework, including the new SC-(e)amfX2C$_{\rm DC}$ models. All energies are given in Hartree.}
 \label{tab:og2-sc-DFT}
 \centering
 \begin{tabular}{l*{5}r}
 \hline \hline\noalign{\smallskip}
 {}&                    {SC-1eX2C$_{\rm D}$} & {SC-\textsf{AMFI}X2C$_{\rm D}$} &    {SC-amfX2C$_{\rm DC}$}&   {SC-eamfX2C$_{\rm DC}$}&  {SC-$^4$DC}\\\hline
$E$&  -109137.69723&  -109137.69723&  -109230.56534&  -109230.56535&  -109230.56535\\
$\epsilon_{1-2}$&    -8210.62133&    -8210.62133&    -8245.93922&    -8245.93921&    -8245.93922\\
$\epsilon_{3-4}$&    -1719.07635&    -1719.07635&    -1725.22140&    -1725.22140&    -1725.22140\\
$\epsilon_{5-10}$&    -1248.58616&    -1248.58616&    -1251.03614&    -1251.03614&    -1251.03614\\
$\epsilon_{11-12}$&     -466.81628&     -466.81628&     -468.19109&     -468.19109&     -468.19109\\
$\epsilon_{13-18}$&     -342.38739&     -342.38739&     -342.98598&     -342.98598&     -342.98598\\
$\epsilon_{19-28}$&     -267.99753&     -267.99753&     -268.22450&     -268.22450&     -268.22450\\
$\epsilon_{29-30}$&     -137.50875&     -137.50875&     -137.87818&     -137.87817&     -137.87817\\
$\epsilon_{31-36}$&      -97.30211&      -97.30211&      -97.45128&      -97.45127&      -97.45127\\
$\epsilon_{37-46}$&      -69.59460&      -69.59460&      -69.63417&      -69.63415&      -69.63415\\
$\epsilon_{47-60}$&      -46.68630&      -46.68630&      -46.68201&      -46.68200&      -46.68200\\
$\ldots$&         $\ldots$&        $\ldots$&       $\ldots$&       $\ldots$&       $\ldots$\\
$\epsilon_{110}$&       -1.29793&       -1.29793&       -1.29619&       -1.29618&       -1.29618\\
$\epsilon_{111}$&       -1.01944&       -1.01944&       -1.02282&       -1.02281&       -1.02281\\
$\epsilon_{112}$&       -1.01877&       -1.01877&       -1.02215&       -1.02215&       -1.02215\\
$\epsilon_{113}$&       -0.28186&       -0.28186&       -0.28190&       -0.28190&       -0.28190\\
$\epsilon_{114}$&       -0.26843&       -0.26843&       -0.26851&       -0.26850&       -0.26850\\
$\epsilon_{115}$&       -0.26843&       -0.26843&       -0.26851&       -0.26850&       -0.26850\\
$\epsilon_{116}$&       -0.26331&       -0.26331&       -0.26339&       -0.26339&       -0.26339\\
$\epsilon_{117}$&       -0.26331&       -0.26331&       -0.26339&       -0.26339&       -0.26339\\
$\epsilon_{118}$&       -0.25140&       -0.25140&       -0.25151&       -0.25150&       -0.25150\\
 \hline \hline
\end{tabular}
\end{table}

\subsubsection{Open-shell Te$_2$}\label{sec:core-orbital-energies-Te2}

In the previous Section \ref{sec:core-orbital-energies-HF}, we primarily focused on the numerical assessment of various 2ePCE corrections schemes for the X2C Hamiltonian in a many-electron context on the basis of the closed-shell superheavy diatomic molecule Og$_2$. In particular, we paid attention to the capability of various 2ePCE-corrected X2C models to provide matching molecular spinor energies by comparison to four-component reference data. In the chemistry of (molecular compounds of) heavy and superheavy elements, one frequently has to cope with partially occupied electronic shells due to the possibility of unfilled $s, p, d$\ and/or $f$\ electronic shells. In order to showcase the versatility of our (e)amf PCE corrections for the X2C Hamiltonian also in such a context, we consider in the following the open-shell molecule Te$_2$. The latter system is a heavy homologue of O$_2$\ and for this reason best characterized by a valence electronic structure that can be written in shorthand as $\pi^4_u \pi^{\ast,2}_g$ (assuming an approximate yet more familiar spin-orbit-free notation of the molecular spinors). For a further, detailed discussion of the electronic structure of the homonuclear diatomic systems of group 16 ranging from O$_2$\ to Po$_2$, we refer the reader, for example, to Ref.~\citenum{rota:114106}. 
As shown in the latter, the molecular bonding ($\pi_u$) and antibonding ($\pi_g^\ast$) combinations predominantly originate from the atomic $p$\ valence shells of each Te atom. Hence, their actual description will be a sensitive measure of an appropriate account of both SC effects and SO coupling. To this end, we will not only consider spin-same-orbit but also spin-other-orbit interaction effects where the latter requires the inclusion of the 2e Gaunt term in the many-body Dirac Hamiltonian.\cite{saue:phd,saue11a}

In Table \ref{tab:te2-hf}, we start our assessment of molecular spinor energies of Te$_2$\ obtained by means of AOC-HF calculations by comparing first data based on various 2ePCE corrections schemes for the X2C Hamiltonian to 4c Dirac-Coulomb Hamiltonian reference values. Notably, for the (closed) core electronic shells we observe for all 2c Hamiltonian schemes similar trends as was the case for Og$_2$ -- with a reference-matching accuracy of our (e)amfX2C models better than $5 \times 10^{-5}$\ Hartree -- which underlines the numerical superiority of our newly proposed PCE correction schemes also in an open-shell case. 
Moving next to the lower end of Table \ref{tab:te2-hf}, that is the (partially) occupied valence $\pi$\ ($\epsilon_{50-51}$) and $\pi^\ast$\ ($\epsilon_{52-53}$) shells, we first note that employing a bare 1eX2C$_D$\ Hamiltonian does not suffice to achieve  sub-mHartree accuraccy in the description of the spin-orbit-split $m_j$\ components of the $\pi^{(\ast)}$\ shells, in particular so for the $\pi^{\ast}_{1/2}-\pi^{\ast}_{3/2}$\ shells ($\epsilon_{52}$\ and $\epsilon_{53}$\ in Table \ref{tab:te2-hf}, respectively). By contrast, -- as opposed to the superheavy diatomic Og$_2$ -- for the heavy Te$_2$\ diatomic system the \textsf{AMFI}X2C$_{\rm D}$\ Hamiltonian yields results for the valence shells on par with the (e)amfX2C$_{\rm DC}$ Hamiltonian both of which are in turn in excellent agreement with the $^4$DC reference.

\begin{table}
\caption{SCF total energy ($E$) and spinor energies ($\epsilon$) of the doubly-degenerate occupied and (partially occupied) open-shell spinors
for Te$_2$ as obtained from AOC/HF/\texttt{v2z} calculations within a four-component Dirac-Coulomb ({$^4$DC}) as well as a two-component Hamiltonian framework, including the new (e)amfX2C$_{\rm DC}$ models. All energies are given in Hartree.}
 \centering
\begin{tabular}{l*{5}r}
\hline \hline \noalign{\smallskip}
{} & {1eX2C$_{\rm D}$} & {\textsf{AMFI}X2C$_{\rm D}$} & {amfX2C$_{\rm DC}$} & {eamfX2C$_{\rm DC}$} & {$^4$DC} \\ \hline
$E$&   -13584.54193&   -13584.34021&   -13587.74121&   -13587.74119&   -13587.74174\\
$\epsilon_{1-2}$&    -1174.97331&    -1174.97784&    -1176.01576&    -1176.01576&    -1176.01572\\
$\epsilon_{3-4}$&     -183.75495&     -183.75645&     -183.87640&     -183.87640&     -183.87640\\
$\epsilon_{5-6}$&     -172.03541&     -171.69323&     -171.76385&     -171.76385&     -171.76385\\
$\epsilon_{7-8}$&     -161.41635&     -161.57069&     -161.63731&     -161.63730&     -161.63731\\
$\epsilon_{9-10}$&     -161.41620&     -161.57054&     -161.63716&     -161.63716&     -161.63716\\
$\epsilon_{11-12}$&      -38.10899&      -38.10952&      -38.13087&      -38.13086&      -38.13086\\
$\epsilon_{13-14}$&      -33.17192&      -33.10276&      -33.11302&      -33.11302&      -33.11302\\
$\epsilon_{15-16}$&      -31.13206&      -31.16370&      -31.17336&      -31.17335&      -31.17335\\
$\epsilon_{17-18}$&      -31.13092&      -31.16256&      -31.17222&      -31.17222&      -31.17222\\
$\epsilon_{19-20}$&      -22.49155&      -22.43146&      -22.43228&      -22.43228&      -22.43228\\
$\epsilon_{21-22}$&      -22.48993&      -22.42984&      -22.43065&      -22.43065&      -22.43065\\
$\epsilon_{23-24}$&      -21.99104&      -22.03063&      -22.03248&      -22.03247&      -22.03247\\
$\epsilon_{25-26}$&      -21.99031&      -22.02990&      -22.03175&      -22.03174&      -22.03174\\
$\epsilon_{27-28}$&      -21.98893&      -22.02852&      -22.03037&      -22.03038&      -22.03038\\
$\ldots$&         $\ldots$&        $\ldots$&       $\ldots$&       $\ldots$&       $\ldots$\\
$\epsilon_{47}$&       -0.86560&       -0.86565&       -0.86590&       -0.86590&       -0.86590\\
$\epsilon_{48}$&       -0.70308&       -0.70312&       -0.70348&       -0.70347&       -0.70347\\
$\epsilon_{49}$&       -0.41423&       -0.41390&       -0.41389&       -0.41389&       -0.41389\\
$\epsilon_{50}$&       -0.36588&       -0.36517&       -0.36514&       -0.36513&       -0.36513\\
$\epsilon_{51}$&       -0.34337&       -0.34389&       -0.34386&       -0.34387&       -0.34387\\
$\epsilon_{52}$&       -0.26021&       -0.25990&       -0.25990&       -0.25990&       -0.25990\\
$\epsilon_{53}$&       -0.23943&       -0.24003&       -0.24003&       -0.24003&       -0.24003\\
\hline \hline
\end{tabular}
\label{tab:te2-hf}
\end{table}

\begin{table}
\caption{SCF total energy ($E$) and spinor energies ($\epsilon$) of occupied spinors for Te$_{2}$ as obtained from Kramers-unrestricted HF/\texttt{v2z} calculations within a four-component Dirac-Coulomb ({$^4$DC}) as well as a two-component Hamiltonian framework, including the new (e)amfX2C$_{\rm DC}$ models. All energies are given in Hartree.}
\centering
\begin{tabular}{l*{4}r}
\hline \hline \noalign{\smallskip}
{} & {1eX2C$_{\rm D}$} & {amfX2C$_{\rm DC}$} & {eamfX2C$_{\rm DC}$} & {$^{4}$DC} \\ \hline
$E$&              -13584.66007&   -13587.85859&   -13587.85862&   -13587.85937   \\
$\epsilon_{1-2}$&  -1174.97217&    -1176.01450&    -1176.01450&    -1176.01466   \\
                &  -1174.96988&    -1176.01219&    -1176.01219&    -1176.01243   \\
$\epsilon_{3-4}$&   -183.75281&     -183.87441&     -183.87441&     -183.87442   \\
                &   -183.75200&     -183.87360&     -183.87360&     -183.87362   \\
$\epsilon_{5-6}$&   -172.03333&     -171.76193&     -171.76192&     -171.76192   \\
                &   -172.03307&     -171.76167&     -171.76166&     -171.76167   \\
$\epsilon_{7-8}$&   -161.41472&     -161.63582&     -161.63581&     -161.63582   \\
                &   -161.41462&     -161.63572&     -161.63571&     -161.63572   \\
$\epsilon_{9-10}$&  -161.41262&     -161.63373&     -161.63373&     -161.63376   \\
                 &  -161.41259&     -161.63370&     -161.63370&     -161.63373   \\
$\epsilon_{11-12}$&  -38.10749&      -38.12953&      -38.12953&      -38.12953   \\
                  &  -38.10495&      -38.12698&      -38.12698&      -38.12698   \\
$\epsilon_{13-14}$&  -33.16988&      -33.11113&      -33.11113&      -33.11113   \\
                  &  -33.16937&      -33.11064&      -33.11063&      -33.11064   \\
$\epsilon_{15-16}$&  -31.13053&      -31.17199&      -31.17199&      -31.17199   \\
                  &  -31.13050&      -31.17195&      -31.17195&      -31.17195   \\
$\epsilon_{17-18}$&  -31.12656&      -31.16803&      -31.16803&      -31.16804   \\
                  &  -31.12652&      -31.16800&      -31.16800&      -31.16800   \\
$\epsilon_{19-20}$&  -22.48891&      -22.42980&      -22.42980&      -22.42980   \\
                  &  -22.48870&      -22.42958&      -22.42958&      -22.42958   \\
$\epsilon_{21-22}$&  -22.48751&      -22.42839&      -22.42838&      -22.42839   \\
                  &  -22.48748&      -22.42836&      -22.42835&      -22.42836   \\
$\epsilon_{23-24}$&  -21.98841&      -22.03001&      -22.03000&      -22.03001   \\
                  &  -21.98819&      -22.02979&      -22.02978&      -22.02978   \\
$\ldots$&             $\ldots$&       $\ldots$&       $\ldots$&       $\ldots$   \\
$\epsilon_{99}$&      -0.41895&       -0.41881&       -0.41881&       -0.41880   \\
$\epsilon_{100}$&     -0.39775&       -0.39742&       -0.39742&       -0.39742   \\
$\epsilon_{101}$&     -0.32364&       -0.32379&       -0.32379&       -0.32379   \\
$\epsilon_{102}$&     -0.32115&       -0.32151&       -0.32151&       -0.32151   \\
$\epsilon_{103}$&     -0.31781&       -0.31769&       -0.31769&       -0.31769   \\
$\epsilon_{104}$&     -0.31521&       -0.31527&       -0.31528&       -0.31528   \\
\hline \hline
\end{tabular}
\label{tab:te2-kuhf}
\end{table}

We note in passing that the excellent  agreement in \textit{absolute} values between \textsf{AMFI}X2C$_{\rm D}$-based data (encompassing spin-same and spin-other-orbit PCE corrections) and the $^4$DCG reference deteriorates not only for the inner-core shells but also for the valence 
$\pi^{(\ast)}$\ manifolds as shown in Table \ref{tab:te2-hf-gaunt}. More importantly, though, \textit{relative} energy differences are, to a large extent, preserved in the valence shells of Te$_2$\ which suggests that the \textsf{AMFI}X2C$_{\rm D}$\ model could still be a viable option for a 2c Hamiltonian framework when aiming for a study of valence-shell dominated molecular properties. Albeit the reasonable relative energy differences in the latter case, to achieve  simultaneously both accurate absolute and relative molecular spinor energies with respect to the $^4$DC as well as $^4$DCG reference data necessitates to resort to our (e)amfX2C Hamiltonian models. As can be inferred from Tables \ref{tab:te2-hf}\ and \ref{tab:te2-hf-gaunt}, both our amf 2ePCE correction models display for \textit{all} electronic shells a numerical accuracy within at least a few $10^{-5}$ Hartree (or better) in comparison to the respective 4c reference.

\begin{table}
\caption{SCF total energy ($E$) and spinor energies ($\epsilon$) of the doubly-degenerate occupied and open-shell spinors
for Te$_2$ as obtained from AOC/HF/\texttt{v2z} calculations within a four-component Dirac-Coulomb-Gaunt ({$^4$DCG}) as well as a two-component Hamiltonian framework, including the new (e)amfX2C$_{\rm DCG}$ models. 
All energies are given in Hartree.}
\centering
\begin{tabular}{l*{5}r}
\hline \hline \noalign{\smallskip}
{} & {1eX2C$_{\rm D}$} & {\textsf{AMFI}X2C$_{\rm D}$} & {amfX2C$_{\rm DCG}$} & {eamfX2C$_{\rm DCG}$} & {$^4$DCG} \\ \hline
$E$&   -13584.54193&   -13584.25457&   -13576.46087&   -13576.46083&   -13576.45740\\
$\epsilon_{1-2}$&    -1174.97331&    -1174.98060&    -1173.19796&    -1173.19796&    -1173.19789\\
$\epsilon_{3-4}$&     -183.75495&     -183.75737&     -183.61571&     -183.61571&     -183.61570\\
$\epsilon_{5-6}$&     -172.03541&     -171.61303&     -171.29409&     -171.29409&     -171.29408\\
$\epsilon_{7-8}$&     -161.41635&     -161.60382&     -161.30279&     -161.30279&     -161.30280\\
$\epsilon_{9-10}$&     -161.41620&     -161.60366&     -161.30263&     -161.30264&     -161.30263\\
$\epsilon_{11-12}$&      -38.10899&      -38.10982&      -38.09468&      -38.09468&      -38.09468\\
$\epsilon_{13-14}$&      -33.17192&      -33.08683&      -33.04133&      -33.04133&      -33.04133\\
$\epsilon_{15-16}$&      -31.13206&      -31.17031&      -31.12709&      -31.12708&      -31.12709\\
$\epsilon_{17-18}$&      -31.13092&      -31.16916&      -31.12594&      -31.12595&      -31.12595\\
$\epsilon_{19-20}$&      -22.49155&      -22.42520&      -22.41206&      -22.41206&      -22.41206\\
$\epsilon_{21-22}$&      -22.48993&      -22.42358&      -22.41044&      -22.41044&      -22.41044\\
$\epsilon_{23-24}$&      -21.99104&      -22.03503&      -22.02348&      -22.02347&      -22.02347\\
$\epsilon_{25-26}$&      -21.99031&      -22.03430&      -22.02275&      -22.02275&      -22.02275\\
$\epsilon_{27-28}$&      -21.98893&      -22.03293&      -22.02138&      -22.02139&      -22.02139\\
$\ldots$&         $\ldots$&        $\ldots$&       $\ldots$&       $\ldots$&       $\ldots$\\
$\epsilon_{47}$&       -0.86560&       -0.86567&       -0.86583&       -0.86582&       -0.86582\\
$\epsilon_{48}$&       -0.70308&       -0.70313&       -0.70323&       -0.70322&       -0.70322\\
$\epsilon_{49}$&       -0.41423&       -0.41382&       -0.41360&       -0.41359&       -0.41360\\
$\epsilon_{50}$&       -0.36588&       -0.36501&       -0.36480&       -0.36479&       -0.36479\\
$\epsilon_{51}$&       -0.34337&       -0.34400&       -0.34382&       -0.34383&       -0.34383\\
$\epsilon_{52}$&       -0.26021&       -0.25983&       -0.25952&       -0.25953&       -0.25953\\
$\epsilon_{53}$&       -0.23943&       -0.24016&       -0.23987&       -0.23987&       -0.23987\\
\hline \hline
\end{tabular}
\label{tab:te2-hf-gaunt}
\end{table}

\subsubsection{Methane -- the ultrarelativistic case}\label{sec:CH4-ultra}

In contrast to the previous molecular examples, methane (CH$_4$)\ consists of a ``heavy" carbon atom C and four ``light" hydrogen atoms H. 
Particularly, since hydrogen is a \textit{one}-electron system, it will not give rise to atomic \textit{two}-electron PCE-correction terms. 
Hence, any genuine \textit{atomic}-mean-field-based PCE-corrected 2c Hamiltonian such as \textsf{AMFI}X2C or amfX2C\ will, by construction, not include any ``light"-atom PCE corrections. 
By contrast, our extended amfX2C approach allows us to eliminate this apparent shortcoming because, as detailed in Section \ref{sec:eamf-and-remarks}\ and outlined in lines \ref{alg:eamfX2C:line4cF}-\ref{alg:eamfX2c:xcenergy}\ of Alg.~\ref{alg:eamfX2C}, all PCE-correction terms for HF and DFT, respectively, are derived in \textit{molecular} basis on the basis of molecular densities, $\vec{D}^{4c}_{\bigoplus}$\ and $\vec{D}^{2c}_{\bigoplus}$, built from a superposition of atomic input densities. Consequently, the essential molecular densities include atomic contributions regardless of the actual atom type, \textit{viz.}\ ``light" (one-electron) and ``heavy" (many-electron) atom contribute on an equal footing.

Bearing the latter in mind, the total SCF energies as well as spinor energies compiled in Table \ref{tab:ch4-DFT}\ for an ultrarelativistic CH$_4$ with the speed of light $c$\ scaled down by a factor 10 confirm the unique numerical performance of the eamfX2C$_{\rm DC}$\ Hamiltonian model in comparison to the $^4$DC\ reference data. Only in the eamfX2C$_{\rm DC}$\ case (column 4, Table \ref{tab:ch4-DFT}), we find that not only the total energy $E$\ agrees to better than mHartree accuracy but also the spinor energies exhibit consistent numerical accuracy for the innermost non-bonding core C $1s$\ as well as the bonding, valence C-H spinors. Notably, the amfX2C$_{\rm DC}$ as well as the \textsf{AMFI}X2C$_{\rm D}$\ models feature an inconsistent numerical performance wrt both quantities: amfX2C$_{\rm DC}$\ yields a total energy $E$\ and spinor energies for the (carbon-centered) inner core spinors $\epsilon_1$\ and $\epsilon_2$, respectively, of the ultrarelativistic CH$_4$\ which are in close agreement with the $^4$DC reference. It shows, however, larger deviations for the valence spinors ($\epsilon_{3-5}$) whereas the opposite conclusions apply to the \textsf{AMFI}X2C$_{\rm D}$-based data. In the latter case, we ascribe the seemingly good performance of the \textsf{AMFI}X2C$_{\rm D}$\ Hamiltonian with errors less than a mHartree in comparison to the $^4$DC reference to a fortuitous error cancellation since the amf based AMFI PCE correction scheme cannot take into account any 2e picture-change corrections that involve contributions from the atomic hydrogen centers. 

\begin{table}
\caption{SCF total energy ($E$) and spinor energies ($\epsilon$) of the doubly-degenerate occupied spinors for CH$_4$ as obtained from DFT/PBE/\texttt{v2z} calculations within a four-component Dirac-Coulomb ({$^4$DC}) as well as a two-component Hamiltonian framework, including the new (e)amfX2C$_{\rm DC}$ models. All energies are given in Hartree. The speed of light $c$\ was reduced by a factor 10.}
 \sisetup{round-mode=places}
 \begin{tabular}{lS[round-precision=5]S[round-precision=5]S[round-precision=5]S[round-precision=5]S[round-precision=5]}
\hline \hline \noalign{\smallskip}
{} & {1eX2C$_{\rm D}$} & {\textsf{AMFI}X2C$_{\rm D}$} & {amfX2C$_{\rm DC}$} & {eamfX2C$_{\rm DC}$} & {$^4$DC} \\ \hline
$E$&  -42.142195949913777&  -42.140385443091688&  -42.264691739896499&  -42.257744533511335&  -42.258497002070968\\
$\epsilon_{1}$  & -10.222060531059&  -10.224014558481&-10.361539605320&-10.360278545853&-10.357940354609\\
$\epsilon_{2}$  & -0.659389228053&   -0.659664901057& -0.662875118124& -0.662689434833& -0.662737426573\\
$\epsilon_{3}$  & -0.357048351154&   -0.352878298679& -0.356492640337& -0.352905557992& -0.352899123441\\
$\epsilon_{4-5}$& -0.333126360981&   -0.335025047337& -0.332817642587& -0.335273918764& -0.335439097817\\
\hline \hline
\end{tabular}
\label{tab:ch4-DFT}
\end{table}

\subsection{Contact densities of copernicium fluorides CnF$_n$}\label{sec:group12-fluorides}

In this section, we assess the accuracy of calculating absolute contact densities as well as the potential to provide reliable relative contact-density shifts computed within PCE-corrected X2C Hamiltonian models by comparing to parent 4c reference data. While absolute contact densities are dominated by contributions of the inner $s$-shells, and to a lesser extent the innermost $p_{1/2}$-shells, of the respective nuclear center of interest, contact-density shifts particularly probe subtle differences of the valence electronic structure and, likewise, polarization of the inner electronic shells both of which originate from the chemical bonding between a reference atom, here the Cn atom, and ligand atoms (or molecules), as, for example, the $n$\ fluorine atoms in the CnF$_n$\ compounds studied in the present work. 
The optimized structures of the CnF$_n$\ ($n=2,4,6$)\ compounds along with the corresponding spatial symmetries are shown in Table \ref{tab:cnfn-structures}. 
Considering the limited basis-set size and point-nucleus approximation in the present work, our optimized Cn-F bond lengths $r_{\rm Cn-F}$ compare reasonably with corresponding benchmark data from a very recent work by 
Hu and Zou \cite{hush21}\ who reported X2C/PBE0-optimized bond lengths $r_{\rm Cn-F}$\ of 1.920, 1.927 and 1.933 \AA\ with an increasing number $n$\ of fluorine ligand atoms. 

\begin{table}
\caption{Four-component DFT/PBE0-optimized structures of CnF$_n$\ ($n=2,4,6$) compounds. For computational details, see text. All internuclear distances $r_{\rm Cn-F}$ are given in \AA.}\label{tab:cnfn-structures}
\centering
 \sisetup{round-mode=places}
 \begin{tabular}{lS[round-precision=4]r}
  \hline \hline\noalign{\smallskip}
    molecule & {$r_{\rm Cn-F}$} & {double group} \\
    & & {symmetry} \\ \hline
     CnF$_2$ & 1.93738700 & D$_{\infty h}^\ast$\\
     CnF$_4$ & 1.9417946469 & C$_{4 h}^\ast$ \\ 
     CnF$_6$ & 1.9477445597 & O$_{h}^\ast$\\ \hline
\end{tabular}
\end{table}

Table \ref{tab:cn-contact}\ summarizes the calculated absolute contact densities as well as density shifts in a spin-dependent (upper panel) and scalar-relativistic (spinfree, lower panel)\ framework. 
As can be seen there, by construction, we find for the bare Cn atom a perfect match for the absolute contact density at the Cn nucleus between our (e)amfX2C$_{\rm DC}$\ PCE-corrected 2c calculations (Table \ref{tab:cn-contact}, entries 4 and 5) and the corresponding 4c reference, irrespective of the inclusion of spin-dependent terms. By contrast, discarding any 2ePCE corrections (1eX2C$_{\rm D}$, entry 2) or including only first-order SO mean-field PCE corrections (\textsf{AMFI}X2C$_{\rm D}$, entry 3) leads to a considerable underestimation of the total contact density. Interestingly, in the \textsf{AMFI}X2C$_{\rm D}$\ case, the total contact density is even smaller than in the 1eX2C$_{\rm D}$\ case and, consequently, in even stronger  disagreement with the 4c reference. Moving next to the difluoride compound, the conclusions surprisingly seem to shift. While all 2c models correctly reproduce the trend of a decrease in the contact density at the Cn nucleus, AMFIX2C$_{\rm D}$\ (923.43 $e/a_0^{3}$, spinfree: 1225.51 $e/a_0^{3}$) now exhibits the best agreement for the contact density shift with the (sc-)$^4$DC reference of 922.84 $e/a_0^{3}$ (1226.40 $e/a_0^{3}$). Considering the remaining tetra- and hexafluoride compounds in Table \ref{tab:cn-contact}, the agreement of \textsf{AMFI}X2C$_{\rm D}$\ for $\Delta\rho$\ with the 4ct references considerably worsens with an increasing number of fluorine ligands. This leads us to conclude that the almost perfect match in $\Delta\rho$\ observed for CnF$_2$\ is likely due to a fortuitous error cancellation. 

What about the (e)amfX2C models? For CnF$_2$, a decomposition of the total contact density at the Cn nucleus in terms of molecular spinor contributions reveals that calculations based on the (e)amfX2C$_{\rm DC}$\ Hamiltonian predict in the spin-dependent case -- similar conclusions hold for the spinfree case -- a major contribution of the Cn $1s$\ shell (\textit{vide supra}) of -43605705.12 $e/a_0^{3}$ (-43605705.33 $e/a_0^{3}$)\ in contrast to the 4c value of -43605699.65 $e/a_0^{3}$. 
Hence, recalling the exact numerical match within significant digits for the bare Cn atom (see Table \ref{tab:cn-contact}, first row), the major source for the difference in the total $\Delta\rho$\ for CnF$_2$\ predominantly traces back to a $\Delta\rho_{1s}\ \approx\ 5.5\ e/a_0^{3}$\ between our 2c (e)amfX2C$_{\rm DC}$\ and the $^4$DC data. 
Moreover, it is precisely for this innermost electronic shell that the molecular spinor energies $\epsilon_{1s}$\ exhibit deviations between (e)amfX2C and $^4$DC on the order of $+3 \times 10^{-4}$\ Hartree. 
In detail, we obtain in both 2c cases $\epsilon_{1s}^{\rm amfX2C_{\rm DC}} = -7117.03293$\ Hartree and $\epsilon_{1s}^{\rm eamfX2C_{\rm DC}} = -7117.03294$ Hartree, respectively, underlining the obvious close relationship of the two approaches, which have to be compared with $\epsilon_{1s}^{^4\rm DC} = -7117.03260$ Hartree.
Despite the slightly increasing discrepancies in $\Delta\rho$\ observed for the remaining polyatomic fluoride compounds of Cn listed in Table \ref{tab:cn-contact}\ which can be explained along the same lines as for the difluoride CnF$_2$\ compound, our (e)amfX2C models yet perform best in a systematic fashion with respect to the four-component references. 
Notably, these encouraging findings hold for both common use cases, with the inclusion of SO interaction and in a genuine spinfree approach. 
In summary, probing the density at a heavy nucleus constitutes an excellent measure of the importance of 2e interaction contributions and, hence, allows us to uniquely reveal even subtle shortcomings of distinct 2ePCE correction models within the X2C Hamiltonian framework by comparing to the corresponding full 4c reference data.

\begin{sidewaystable}
  \caption{Contact densities $\rho$\ and contact density shifts $\Delta\rho$\ evaluated at the Cn nucleus for the Cn atom and different Cn fluoride compounds. All data was obtained from scalar-relativistic+spin-orbit (upper panel) and scalar-relativistic-only spinfree (lower panel) HF wave functions. For the two-component X2C Hamiltonian different two-electron picture-change effect corrections were employed, including the new (e)amfX2C$_{\rm DC}$ models. All densities are given in $e/a_0^{3}$.}
 \centering
 \sisetup{round-mode=places}
 \begin{tabular}{lS[round-precision=2]S[round-precision=2]S[round-precision=2]S[round-precision=2]S[round-precision=2]}
  \hline \hline\noalign{\smallskip}
{compound}     & {1eX2C$_{\rm D}$} & {\textsf{AMFI}X2C$_{\rm D}$}       & {amfX2C$_{\rm DC}$} & {eamfX2C$_{\rm DC}$}      & {$^{4}$DC}     \\ \hline
Cn             & -58697556.08172  & -58661390.25980  & -58977494.38640 & -58977494.38640 & -58977494.38620 \\
CnF$_2$        & -58696660.51485  & -58660466.83     & -58976578.21    & -58976577.93992 & -58976571.54    \\  
CnF$_4$        & -58695900.03473  & -58659721.50     & -58975824.80    & -58975823.86983 & -58975812.06    \\  
CnF$_6$        & -58695683.20503  & -58659514.86     & -58975609.89    & -58975608.49949 & -58975593.75    \\  
$\Delta\rho_{(\rm CnF_2-Cn)}$ & 895.5668700039387  &   923.43 &  916.18 &  916.446479998529  &  922.84  \\  
$\Delta\rho_{(\rm CnF_4-Cn)}$ & 1656.0469899997115 &  1668.76 & 1669.58 & 1670.5165700018406 & 1682.33  \\  
$\Delta\rho_{(\rm CnF_6-Cn)}$ & 1872.8766900002956 &  1875.40 & 1884.50 & 1885.8869099989533 & 1900.64  \\ \hline
    \multicolumn{6}{c}{        }             \\[-2ex]
    \multicolumn{6}{c}{spinfree}             \\[0.5ex]
Cn                   &  -56251080.52956 & -56251080.53    & -56571626.66    & -56571626.66086 &  -56571626.66    \\  
CnF$_2$              &  -56249855.02    & -56249855.02    & -56570406.42    & -56570405.90180 &  -56570400.26    \\  
CnF$_4$              &  -56249040.54420 & -56249040.54420 & -56569579.68666 & -56569578.93171 &  -56569569.38740 \\
CnF$_6$              &  -56248766.28    & -56248766.28    & -56569295.28    & -56569294.35461 &  -56569283.33    \\  
$\Delta\rho_{(\rm CnF_2-Cn)}$ & 1225.51            & 1225.51            & 1219.24             & 1220.7590600028634 & 1226.40 \\
$\Delta\rho_{(\rm CnF_4-Cn)}$ & 2039.9853599965572 & 2039.9853599965572 & 2046.973339997232   & 2047.729150004685  & 2057.2725999951363 \\
$\Delta\rho_{(\rm CnF_6-Cn)}$ & 2314.25            & 2314.25            & 2331.38             & 2332.30624999851   & 2343.33 \\ \hline

\label{tab:cn-contact}
\end{tabular}
\end{sidewaystable}

\subsection{X-ray core ionization energies}\label{sec:group17-anion}

Finally, we compare the performance and reliability of the 1eX2C, AMFIX2C as well as (e)amfX2C 2c Hamiltonian models for the calculation of X-ray core ionization energies by comparing to corresponding mmfX2C reference values. 
With the advent and general accessibility of new, powerful X-ray radiation sources such as free-electron lasers \cite{pell16a} (see for example Ref.~\citenum{xefl}\ for an overview of available facilities), experimental X-ray spectroscopies have witnessed in the past decade a continuous, rapid advance and enhanced applicability to study not only the electronic structure but also the dynamics of molecules and materials.\cite{linf17a,cher17a,krau18a} In order to keep pace with the experimental progress and being able to provide a much welcomed highly accurate theoretical support, computational X-ray spectroscopy has experienced tremendous progress in recent years.\cite{norm18a} Here, a genuine inclusion of relativistic effects is nothing but a basic requirement since the inner-core shells are most prone to quantitative changes due to relativity. For example, while K-edge X-ray spectroscopy probes the chemical nature of the 1$s_{1/2}$\ shell of a given center\ and, hence, necessitates in particular a proper account of SC contributions, studying the L- and M-edge of (late) transition-metal, $p$-block\ and, perhaps most importantly, $f$-elements \cite{kvas21a}, whose fine-structure is dominated by the SO splitting of the 2$p$- and $3p$- and 3$d$-shells, respectively, requires a suitable framework to efficiently take into account the SO interaction. The latter two requirements are easily met in either a (exact) 2c or full 4c framework that sets out from a many-particle Dirac-Coulomb(-Gaunt/-Breit) Hamiltonian. For further details and recent advances of genuine relativistic quantum-chemical X-ray spectrocsopy approaches that illustrate in a striking fashion the potential of such ans{\"a}tze, we refer the reader, for example, to Refs.~\citenum{Kadek2015,saue:uraniumX2015,stet19a,halb21,kone22a}.

\begin{table}
  \caption{EOM-CCSD/\texttt{dyall.acv3z} core-ionization energies of the At$^-$ anion obtained within a two-component Hamiltonian framework employing different corrections for two-electron picture-change effects. Note, that for At$^-$\ amfX2C$_{\rm DC}$ and eamfX2C$_{\rm DC}$\ yield identical results. All energies are given in Hartree.}
 \centering
 \sisetup{round-mode=places}
 \begin{tabular}{lS[round-precision=4]S[round-precision=4]S[round-precision=4]S[round-precision=4]S[round-precision=4]}
  \hline \hline\noalign{\smallskip}
  Ionization & {1eX2C$_{\rm D}$} & {\textsf{AMFI}X2C$_{\rm D}$} &
   {amfX2C$_{\rm DC}$} & {amfX2C$^a_{\rm DC}$}& {mmfX2C$_{\rm DC}$\ $^b$}  \\ \hline
  K-edge & 3532.894929650263 & 3532.939317909214 & 3538.264036722830  & 3538.264186757132 &   3538.263866922428\\ 
  L$_1$-edge & 644.591254197807 & 644.605932617072  & 645.429009582471 & 645.429031765865 &   645.428978040168\\ 
  L$_2$-edge & 620.862473993097 & 618.761888475248 & 619.273038616374 &   619.273046689022 & 619.272819348205\\ 
  L$_3$-edge & 522.513701785800 & 523.296766499242 & 523.709240319375 & 523.709244256276 &  523.709235139483\\ \hline
$\Delta_{\rm L}^{\rm SO}$ & 98.34877220729697 & 95.465121976006 & 95.56379829699904 & 95.563802432746 & 95.56358420872198\\ \hline
  \multicolumn{6}{l}{$^a$\ amf corrections calculated for a neutral At atom.}\\
  \multicolumn{6}{l}{$^b$\ {mmfX2C}$_{\rm DC} \equiv\ ^2$DC$^\textsf{m}$\ values taken from Ref.~\citenum{halb21}.}
    \end{tabular}
\label{tab:group-17-At-}
\end{table}

Considering common applications in X-ray spectroscopy, we highlight in Tables \ref{tab:group-17-At-}\ and \ref{tab:mol-IPs} the importance of 2ePCE corrections to the X2C Hamiltonian which we may anticipate, based on all findings discussed in the previous sections (\textit{vide supra}), to be most pronounced for the K- up to M-edges of heavy- and superheavy nuclei. 
Starting with the EOM-CCSD core-ionization potentials of the heavy $p$-block anion At$^-$\ compiled in Table \ref{tab:group-17-At-}, we note that the K-edge ionisation potentials for within the 1eX2C$_{\rm D}$\ and \textsf{AMFI}X2C${\rm D}$\ Hamiltonian frameworks deviate more than 5 Hartree (sic!) from the mmfX2C$_{\rm DC}$ reference. Concerning the use of the latter, it was shown in Ref.~\citenum{halb21}\ that making use of this 2c Hamiltonian scheme yields ionization potentials which are virtually indistinguishable from the parent $^4$DC data and this is indeed confirmed by the present calculations.  
Moving to our (e)amfX2C PCE-corrected Hamiltonian framework, we observe an agreement with the mmfX2C$_{\rm DC}$ data of sub-mHartree accuracy not only for the K- but also for the L$_1$\ as well as L$_{2,3}$\ edges. 
The resulting deviation of 27 cm$^{-1}$\ from the reference data for the SO-splitting $\Delta_{\rm L-edge}^{\rm SO}$\ (5th row, Table \ref{tab:group-17-At-}), that ultimately governs the fine-structure of the L$_{2,3}$\ edges, approaches almost \textit{spectroscopic} accuracy of 1 cm$^{-1}$ \cite{puzz19a}. 
By contrast, the error for $\Delta_{\rm L-edge}^{\rm SO}$\ in the case of employing, for example, the hitherto popular \textsf{AMFI}X2C$_{\rm D}$\ Hamiltonian is as large as 21600 cm$^{-1}$ (corresponding to an error that is 60 times (sic!) larger than the error bar for \textit{chemical} accuracy).

\begin{table}
  \caption{EOM-CCSD/\texttt{v2z} core-ionization energies of the molecular compounds [Au(Cl)$_4$]$^-$ and CnF$_6$\ obtained within a two-component Hamiltonian framework employing different corrections for two-electron picture-change effects, including the new (e)amfX2C$_{\rm DC}$ models. All energies are given in Hartree.}
 \centering
 \sisetup{round-mode=places}
 \begin{tabular}{lS[round-precision=4]S[round-precision=4]S[round-precision=4]S[round-precision=4]S[round-precision=4]S[round-precision=4]}
%
%
\hline \hline\noalign{\smallskip} Ionization & {1eX2C$_{\rm D}$} & {\textsf{AMFI}X2C$_{\rm D}$} & {amfX2C$_{\rm DC}$} & {eamfX2C$_{\rm DC}$} &  {mmfX2C$_{\rm DC}$}  & {$^4$DC}\\ \hline
\multicolumn{7}{c}{}\\[-2.0ex]
\multicolumn{7}{c}{[AuCl$_4$]$^-$}\\[0.5ex]
    K-edge                & 2982.830463559245 & {n/a} & 2986.9702 & 2986.970218271871 &  2986.9702 & 2986.970498065536 \\
L$_1$-edge                &  531.126692640790 & {n/a} & 531.738635527489 & 531.738652849833 &  531.738573729483 & 531.7387360736\\
L$_2$-edge                &  509.782327229424 & {n/a} & 508.581018886277  & 508.581036312093 &  508.580898693103 & 508.580751385353 \\
L$_3$-edge                &  440.079408044323 & {n/a} &  441.006049899916 &  441.006060238183 &  441.006006416050 & 441.006421957590 \\
                          & 440.079228506535  & {n/a} & 441.005767619948 & 441.005790369592 & 441.005754701790 & 
                          441.006149449236 \\
M$_4$-edge                & 85.643053043861 & {n/a} &
                            85.348628234523 &           85.348652944471 &           85.348645688948 &
                            85.348527314125\\
                          & 85.641399162724 & {n/a} &    85.34711289382  &            85.347128962702 &            85.347125715726 & 
                            85.347004152568\\
M$_5$-edge                & 81.911683458561 & {n/a} &                              82.132262178190 &                                      82.132279211819 &                                      82.132245139735 &                                      82.132457493129\\ 
                          & 81.909670637807 & {n/a} & 82.130240618601 & 82.130268761688 & 82.130244579613 &  82.130456191236\\
                          & 81.908228588158 & {n/a} & 82.128969359407 & 82.128979219543 & 82.128984659749 & 82.129184389708  \\ \hline \hline
$^a\Delta_{\rm L-edge}^{\rm SO}$ &  69.70300895399504 & {n/a} & 67.57511012634495 & 67.57511100820551 & 67.57501813418298 & 67.57446568193996 \\ 
$\Delta \Delta_{\rm L-edge}^{\rm SO}$ & 2.1280 & {n/a} & 0.0001 & 0.0001 & {0} & {-} \\ \hline
%
%
\multicolumn{7}{c}{}\\[-2.0ex]
\multicolumn{7}{c}{CnF$_6$}\\[0.5ex]
    K-edge                & 7098.864222757124 & 7099.085622799346 & 7116.459673602369 & 7116.459717478101 & 7116.458951391109 & 7116.458479846372 \\
L$_1$-edge                & 1450.835145507292 & 1450.906556831415 & 1454.398298451817 & 1454.398350552363 & 1454.397944362923 & 1454.398304520716 \\
L$_2$-edge                & 1412.330779914150 & 1404.951102695315 & 1406.920094078070 & 1406.920144490339 & 1406.919752354087 & 1406.919409388566 \\
L$_3$-edge                & 1003.154441459877 & 1005.198622756358 & 1006.439514781305 & 1006.439567474321 & 1006.439338856743 & 1006.440092360885 \\ \hline \hline
$\Delta_{\rm L-edge}^{\rm SO}$ & 409.17633845427304& 399.752479939  & 400.4805792968 & 400.480577016 & 400.4804134973  & 400.47931702768096 \\
$\Delta \Delta_{\rm L-edge}^{\rm SO}$ & 8.695924956973045 & -0.7279335582999806  & 0.00016579949999595556 & 0.00016351869999198243 & {0}  & {-} \\
\hline
\multicolumn{7}{l}{$^a$\ calculated as $\Delta^{\rm SO}$(L$_2$-$\bar{\rm L}_3$)\ using an arithmetic mean value for the L$_3$-edge.}
    \end{tabular}
\label{tab:mol-IPs}
\end{table}







Table \ref{tab:mol-IPs}\ compiles core-ionization potential data for two representative molecular $5d$- (upper panel) and $6d$\ (lower panel) complexes as obtained from EOM-CCSD calculations. As was the case for the At$^-$\ anion, we consider the numerical performance of different \textit{atomic} mean-field 2ePCE-correction schemes for the X2C Hamiltonian by comparing to results calculated within a \textit{molecular} mean-field 2c framework (Table \ref{tab:mol-IPs}, entry 6). In passing we note that for the [Au]-complex (upper panel of Table \ref{tab:mol-IPs}), we were not able to obtain a converged SCF solution for Au within the external SCF program \textsc{relscf} \cite{relscf} that constitutes the basis for the AMFI module within \textsc{DIRAC}, and this is unfortunately a recurring problem. Considering first the full neglect of 2ePCE corrections within the 1eX2C$_{\rm D}$\ framework (Table \ref{tab:mol-IPs}, entry 2), a similar picture emerges in both molecular cases as in the single-ion case. The absolute deviations for the ionization potentials of all K- to M-edges are substantial. Moreover, the same conclusions hold for relative deviations, exemplified by the SO-splittings $\Delta_{\rm L-edge}^{\rm SO}$\ of the L-edge. 
Hence, these findings unequivocally demonstrate also in the context of X-ray spectroscopic quantities that 2ePCEs are substantial when  probing molecular properties of the inner-core shells. 
Interestingly, though, the ligand-field induced splittings of the M$_{4,5}$-edges in the case of the [Au]-complex can be correctly reproduced within the 1eX2C$_{\rm D}$\ Hamiltonian framework. 
As can be seen for the CnF$_2$\ complex, the inclusion of first-order mean-field SO PCE corrections (entry 3, Table \ref{tab:mol-IPs}) within the \textsf{AMFI}X2C$_{\rm D}$\ Hamiltonian leads to a reduction of the error for $\Delta_{\rm L-edge}^{\rm SO}$\ by one order of magnitude\ from $\Delta \Delta_{\rm L-edge}^{\rm SO} \approx +8.7$\ Hartree (1eX2C$_{\rm D}$) to $\Delta \Delta_{\rm L-edge}^{\rm SO} \approx-0.7$\ Hartree. Still, the underlying absolute core-ionization potentials 
for the K- and L-edges exhibit a clear deviation ranging from approximately 1.2 Hartree for the L$_3$-edge to more than 17 Hartree for the K-edge in comparison to the mmfX2C$_{\rm DC}$ data. 

By contrast, the EOM-CCSD core-ionization potentials calculated within the (e)amfX2C$_{\rm DC}$\ Hamiltonian frameworks (entries 4 and 5 in Table \ref{tab:mol-IPs}) stand out also in the molecular cases due to two distinct, appealing features, namely (i) the \textit{absolute} ionization energies for all edges feature numerical values below sub-mHartree accuracy and (ii), as a result, this accuracy carries over to relative data such as the SO splitting of the L$_{2,3}$-edge\ and the ligand-field fine-structure splitting of the M$_{4,5}$-edges in the [Au]-complex. Hence, the atomic-meanfield (e)amfX2C Hamiltonian models can be regarded as a conceptually different alternative to the molecular mean-field $^2$DC scheme by providing  virtually the same numerical accuracy for core- and likewise valence molecular properties at a fraction of the computational effort. To stress the latter, we recall that the mmfX2C$_{\rm DC}$\  approach requires to first find a converged  \textit{molecular} 4c SCF solution whereas our (e)amfX2C models are solely built on quantities obtained from \textit{atomic}\ 4c SCF calculations. In the latter case, the SCF step is then carried out exclusively in a molecular 2c framework. Moreover, we note that, although the extended amfX2C Hamiltonian model requires the calculation of a single 2e Fock matrix $\vec{F}^{4c,2e}[\vec{D}_{\bigoplus}^{4c}]$\ in a molecular four-component framework, an efficient density-matrix-based screening will significantly reduce the associated computational cost because of the sparsity of the atom-wise blocked 4c molecular density matrix $\vec{D}_{\bigoplus}^{4c}$. 

\begin{figure}
    \centering
    \includegraphics[scale=0.75]{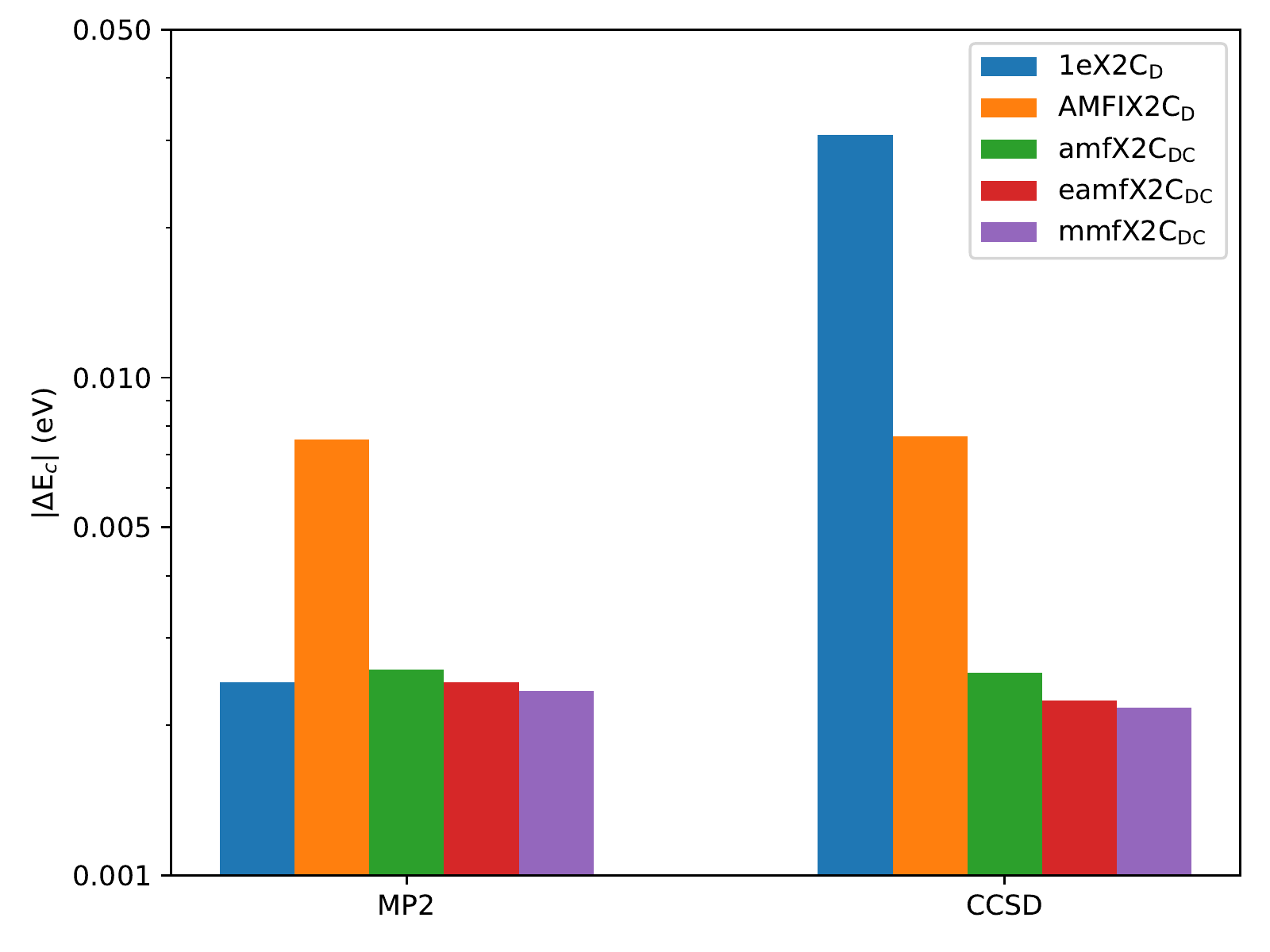}
    \caption{Absolute MP2 and CCSD correlation energies differences between $^4$DC ($\left|\Delta E_c\right|$, in eV), for different Hamiltonian calculated for CnF$_6$\ with the same computational setup as for the EOM-CCSD core-ionization energies. Note that the scale on the y-axis is logarithmic. Further computational details are given in the text.}
    \label{fig:cnf6-corr-e}
\end{figure}

 Besides the calculation of core-binding energies within a 2c Hamiltonian framework taking into account various PCE correction models, we performed $^4$DC-based EOM-CCSD calculation for the [Au]-complex as well as CnF$_6$\ (Table \ref{tab:mol-IPs}, entry 7). This enables us to further assess the inﬂuence of the Hamiltonian on the core ionizations, in particular inherent PCEs in the electron-electron interaction within a two-component X2C framework, regardless of an mmf or amf model to account for PCEs. As discussed in detail by Halbert \textit{et al.}\ in Ref.~\citenum{halb21}\ in the context of X-ray core binding energies, even for mmfX2C$_{\rm DC}$, which is based on the transformation into a 2c framework wrt a decoupling of the (converged) Fock matrix, leaving the 2e operator untransformed \cite{sikkema09}\ necessarily introduces a PCE in the electron-electron interaction. Hence, the latter becomes most prominent for molecular properties that necessitate an accurate treatment of core-core and core-valence electron correlation such as X-ray core ionization potentials. 
 Consequently, albeit our limited correlation treatment in the EOM-CCSD step (see Section \ref{sec:compdet}\ for further details), we already find small discrepancies for the K- and L-edge ionization energies between mmfX2C$_{\rm DC}$ and $^4$DC on the order of 0.0025 eV and, similarly, for (e)amfX2C and $^4$DC with differences up to 0.0035 eV in the case of CnF$_6$\ while the deviations in the binding energies for the corresponding edges are smaller for the [Au]-complex because of the ``lighter" Au central atom. As larger deviations -- though still less than 0.01\% of the total K-edge  binding energy -- had been observed in a corresponding comparison for astatine \cite{halb21}, we expect also for CnF$_6$\ (and, similarly, for the [Au]-complex) a further increase of the deviations between $^4$DC\ and (e)amfX2C$_{\rm DC}$ as well as mmfX2C$_{\rm DC}$\ upon an improved electron-correlation treatment. In passing we note, though, the excellent performance of our extended amfX2C-based computational model (red error bars in Figure \ref{fig:cnf6-corr-e}) with respect to the mmfX2C$_{\rm DC}$ model (purple error bars in Figure \ref{fig:cnf6-corr-e})\ which are nearly identical for both MP2\ and CCSD correlation energies even on a logarithmic scale in the case of CnF$_6$. By contrast, turning to 1eX2C$_{\rm D}$\ and \textsf{AMFI}X2C$_{\rm D}$, respectively, we either find stark differences in the correlation errors between the MP2 and CCSD approaches (1eX2C$_{\rm D}$)\ or, even within this limited correlation space, considerable errors in the correlation energies within \textsf{AMFI}X2C$_{\rm D}$\ by comparison to the ones obtained within a $^4$DC\ framework. Taking together, these findings yet again underline the suitability and superiority of our amfX2C Hamiltonian models, in particular its extended variant, in the realm of an X2C framework for studying X-ray core binding energies of atoms and molecules comprising heavy and superheavy elements.

\begin{table}
  \caption{EOM-CCSD/\texttt{v2z} core-ionization energies of CnF$_6$\ including contributions from the two-electron Gaunt interaction obtained within a two-component Hamiltonian framework employing different corrections for two-electron picture-change effects, including the new (e)amfX2C$_{\rm DCG}$ models. All energies are given in Hartree.}
 \centering
 \sisetup{round-mode=places}
 \begin{tabular}{lS[round-precision=4]S[round-precision=4]S[round-precision=4]S[round-precision=4]S[round-precision=4]}
%
%
\hline \hline\noalign{\smallskip} Ionization & {1eX2C$_{\rm D}$} & {\textsf{AMFI}X2C$_{\rm D}$} & {amfX2C$_{\rm DCG}$} & {eamfX2C$_{\rm DCG}$} &  {mmfX2C$_{\rm DCG}$}  \\ \hline
\multicolumn{6}{c}{}\\[-2.0ex]
    K-edge                & 7098.864222757124 & 7099.244902278329 & 7076.503732188685 & 7076.503830421792 & 7076.501403453378\\
L$_1$-edge                & 1450.835145507292 & 1450.960331295658 & 1448.800870735362 & 1448.800972562912 & 1448.800310220434 \\
L$_2$-edge                & 1412.330779914150 & 1402.780910028746 & 1397.811070079591 & 1397.811170997760 & 1397.810400069455 \\
L$_3$-edge                & 1003.154441459877 & 1005.551885385839 & 1001.540331728849 & 1001.540433936657 & 1001.540508582275 \\ \hline \hline
$\Delta_{\rm L-edge}^{\rm SO}$ & 409.17633845427304& 397.2290246429069  & 396.270738350742 & 396.270737061103 & 396.26989148717996  \\
$\Delta \Delta_{\rm L-edge}^{\rm SO}$ & 12.906446967093075 & 0.9591331557269314  & 0.0008468635620602072 & 0.0008455739230157633 & {0}  \\
\hline
    \end{tabular}
\label{tab:mol-IPs-Gaunt}
\end{table}

Finally, Table \ref{tab:mol-IPs-Gaunt}\ compiles X-ray EOM-CSCD core binding energies of CnF$_6$\ with the inclusion of the 2e Gaunt interaction. This allows us to highlight the significance of the Gaunt interaction (as part of the full Breit interaction) for an accurate description of the inner-core edges of (super-)heavy elements by comparing to the corresponding Coulomb-type-interaction only data shown above in Table \ref{tab:mol-IPs}. Note that we do not have $^{4}$DCG data at hand since the transformation of Gaunt-type AO integrals to MO basis is currently neither implemented in \textsc{DIRAC}\ nor in \textsc{ReSpect}. In addition, as discussed for Te$_2$\ in Section \ref{sec:core-orbital-energies-Te2}, 1eX2C$_{\rm D}$\ does not allow to take into account\ contributions from the Gaunt interaction and will not be considered further below. 

In agreement to what has been concluded in Ref.~\citenum{halb21}\ for astatide, we find for the (e)amfX2C$_{\rm DCG}$ as well as mmfX2C$_{\rm DCG}$ models (Table \ref{tab:mol-IPs-Gaunt}, entries 4-6) a distinct effect arising from the Gaunt interaction. As a result, core-binding energies are substantially lowered by nearly 40 Hartree ($\approx$ 1.1 keV (!)) in case of the K- and by up to 9 Hartree for the L-edge, respectively. Moreover, we also observe a considerable decrease in the SO splitting of the L$_{2,3}$-edge $\Delta_{\rm L-edge}^{\rm SO}$\ by approximately 115 eV which is very well captured ($\Delta \Delta_{\rm L-edge}^{\rm SO} \approx 0.02$ eV) by our (e)amfX2C$_{\rm DCG}$ models in comparison to mmfX2C$_{\rm DCG}$. The latter is in sharp contrast to the \textsf{AMFI}X2C$_{\rm D}$\ model (entry 3) which not only exhibits significant numerical differences in terms of absolute core-binding energies of more than 600 eV for the K-edge but also shows a quantitative error $\Delta \Delta_{\rm L-edge}^{\rm SO}$\ of more than 25 eV for the L$_{2,3}$-edge fine-structure splitting. 

\section{Conclusions and Outlook}\label{sec:conclu}

In this article we have presented the motivation for and derivation of two distinct, atomic-mean-field (amf) approaches to account on an equal footing for two-electron (2e) scalar-relativistic and spin-orbit picture-change effects (PCEs) arising within an exact-two-component (X2C) Hamiltonian framework. 
Both approaches, dubbed amfX2C and extended amf (eamfX2C) have been implemented independently in the \textsc{Dirac}~\cite{dirac-paper} and \textsc{ReSpect}~\cite{ReSpect2020}  programs.
These implementations, which exploit -- where available -- atomic supersymmetry in the atomic self-consistent field steps,\cite{dirac-paper} open up for the calculation of two-electron picture-change effect corrections for all spin-dependent and spinfree four-component-based Hamiltonians available in the two quantum-chemical software packages. 

Notably, we have shown that it is possible to uniquely tailor our amf 2ePCE corrections for the X2C Hamiltonian to the underlying classes of self-consistent field (SCF) \textit{ans\"atze}, namely Hartree-Fock (HF) or density functional theory (DFT). 
Such a particular feature has, to the best of our knowledge, so far not been considered for any 2ePCE correction scheme in the literature. 
Moreover, by contrast to, for example, the recently proposed SOAMFX2C model\cite{liuj18} our new PCE correction schemes for the X2C Hamiltonian take into account both \textit{spin-independent} -- that is scalar-relativistic -- and \textit{spin-dependent} -- that is spin-spin (arising from the Gaunt term\cite{saue:phd}) as well as spin-orbit -- contributions of the two-electron interaction. 
Perhaps most importantly, we also argue why the eamfX2C Hamiltonian can be employed in genuine two-component solid-state SCF calculations under consideration of periodic boundary conditions\ starting from an appropriate four-component framework.\cite{kade19} 
The latter is subject of ongoing work in our laboratories.

The novel (e)amfX2C models are readily available for genuine  two-component atomic and molecular SCF calculations including both HF and DFT. As these then often constitute the basis for more elaborate approaches such as (real-time) time-dependent \textit{ans\"atze} as well as response-theory based approaches and post-HF electron-correlation approaches in general, for example, configuration-interaction- or coupled-cluster-type wave-function expansions, our (e)amfX2C models are broadly applicable within a two-component quantum-chemical framework.  

As a first demonstration of the capabilities of the (e)amfX2C models, we have applied them to the calculation of molecular spinor energies of representative closed and open-shell (super-)heavy homonuclear diatomic molecules of group 16 and 18, respectively, both within an HF and a DFT-based SCF \textit{ansatz}. 
With these systems, namely Te$_2$\ and Og$_2$, we have assessed the numerical accuracy of the (e)amfX2C Hamiltonian models by comparing to four-component reference data with respect to the ability to reproduce absolute spinor energies as well as relative energies defined as the atomic-like spin-orbit splittings of the inner-core shells. 
As a further test, we have calculated both the absolute contact density at the Cn nucleus and contact density shifts in copernicium fluoride compounds (CnF$_n$, $n=2,4,6$) relative to the atomic value for the bare Cn atom. 
Finally, we have studied the performance of our (e)amfX2C Hamiltonian models for core-electron binding energies in the realm of X-ray spectroscopy by making use of an equation-of-motion coupled-cluster approach. 

For the open- and closed-shell diatomic molecules we demonstrate that by applying our (e)amf PCE corrections to the X2C Hamiltonian models it is possible to match all corresponding four-component molecular spinor energies with $\mu$-Hartree accuracy, \textit{viz.} for inner- core to outer-valence electronic shells. 
This outstanding performance holds not only for two-component SCF calculations within a Kramers-restricted and Kramers-unrestricted HF \textit{ansatz} but also within a DFT framework. 
Moreover, we show that scalar-relativistic two-electron PCE corrections are of utmost importance for a reliable description of core electronic shells within a two-component X2C Hamiltonian framework. 
The latter necessity manifests itself also in the calculation of absolute as well as relative contact densities at the Cn nucleus with respect to CnF$_n$\ ($n=0,2,4,6$)\ compounds, where their neglect can lead to sizeable discrepancies with respect to the same quantities obtained within a four-component framework. 
Although the (e)amf corrections are able to eliminate a substantial part of the scalar-relativistic and spin-orbit two-electron PCEs in the X2C framework, qualitative discrepancies between our two- and four-component results remain. We could trace the missing gap to the $^4$DC reference data for the absolute contact density at the Cn nucleus in CnF$_2$\ and, similarly, for the other CnF$_n$\ ($n>2$) compounds, to originate from a contact density contribution of the Cn $1s$\ shell whose contributions show a relative deviation of about 11\%\ between two- and four-component data. 

A comparison of X-ray core binding energies for At$^{-}$, [AuCl$_4$]$^{-1}$\ and CnF$_6$\ further highlights the significance of an appropriate account of two-electron PCE corrections in a two-component framework in order to unambiguously and systemically improvable 
approach $^4$DC(G) results. In particular, we demonstrate that 
our (e)amfX2C models enable X2C calculations of X-ray ionisation potentials  -- and the accompanying resolution of fine-structure fingerprints of L- and M-edges in heavy- and superheavy-element complexes -- where the transformation to two-components is performed \textit{prior} to the (molecular) SCF step while yielding results both on par with corresponding \textit{molecular} mean-field calculations and in excellent agreement with the parent four-component ones. Moreover, we illustrate that it is possible within our (e)amfX2C models to account for two-electron effects originating from the Gaunt interaction. To ultimately strive for genuine comparisons of computed X-ray spectroscopic data with experiment, an inclusion of the full Breit interaction, higher-order correlation effects as well as QED-corrections will be essential to establish a computational model of true predictive power.\cite{halb21} While the former two factors are currently under consideration within the \textsc{DIRAC}\ developers community ,\cite{dirac-meeting2021} QED corrections have very recently been put forward for correlated calculations in a two-component framework\cite{suna21a} and will be made available in a future extension of our (e)amfX2C models.

In summary, we are confident that the picture-change-error correction models for the X2C Hamiltonian presented in this contribution constitute an important milestone towards a universal and reliable applicability of relativistic two-component quantum chemical approaches maintaining the accuracy of the parent four-component one at a fraction of its computational cost. 
In order to corroborate the latter, we are currently undertaking comprehensive studies of zero-field splittings in $p$- and $d$-block molecules as well as the calculations of EPR parameters of $d$- and $f$-element complexes on the basis of our (e)amfX2C Hamiltonian models within a correlated computational framework. Finally, since relativistic real-time time-dependent DFT \cite{repi15a} and wave-function based correlated approaches such as the density matrix renormalization group model \cite{baia22a} provide access to the absorption spectra of complex molecular systems in the valence- or core-excited range including a variational account of spin-orbit interaction, we intend to apply these approaches within our (e)amfX2C framework to a set of representative molecular $d$-block and actinide compounds. 


\section*{Dedication}
We dedicate this work to the memory of the late Bernd Schimmelpfennig, who
passed away unexpectedly in 2019. He was, among other contributions, a pioneer
in making corrections for two-electron picture-change effects within a
two-component Hamiltonian framework not only popular but also, for the first
time, widely usable in quantum chemistry.

\begin{acknowledgments}
SK thanks the SHC department at GSI Darmstadt for continuing support. Part of the calculations were performed at the High-Performance Computing infrastructure provided by the GSI IT Department.
MR acknowledges supports from the Research Council of Norway through a Centre of Excellence Grant (Grant No 252569) and a research grant (Grant No 315822), as well as computational resources provided by UNINETT Sigma2 -- the National Infrastructure for High Performance Computing and Data Storage in Norway (Grant No. NN4654K).
TS acknowledges funding from the European Research Council (ERC) under the European Union's Horizon 2020 research and innovation programme (grant agreement No 101019907).
\end{acknowledgments}

\section*{Author's contributions}
All authors contributed equally to theory development. All programming and calculations were carried out by SK (\textsc{DIRAC}) and MR (\textsc{ReSpect}). 

\section*{Data availability}\label{sec:SIData}
The data that support the findings of this study are openly available in ZENODO at \url{https://doi.org/10.5281/zenodo.6414910}, see also Ref.~\citenum{ReplicationData}.

\section*{Conflicts of interest}
The authors have no conflicts to disclose.


\bibliography{amfX2C}


\end{document}